\documentclass[11pt,english]{article}
\usepackage[latin9]{inputenc}
\usepackage{amsmath}
\usepackage{graphicx}
\usepackage{setspace}
\makeatletter

\providecommand{\tabularnewline}{\\}

\@ifundefined{date}{}{\date{}}
\usepackage{esint}
\usepackage{hyperref}
\usepackage{latexsym}
\usepackage{graphicx}\usepackage{bm}\usepackage{longtable}

\usepackage{xcolor}

\@addtoreset{equation}{section}

\makeatother

\usepackage{babel}
\begin{document}
\title{Corrections to the instanton configuration as baryon in holographic
QCD}
\maketitle
\begin{center}
Si-wen Li\footnote{Email: siwenli@dlmu.edu.cn}, Hao-qian Li\footnote{Email: lihaoqian@dlmu.edu.cn},
Sen-kai Luo\footnote{Email: luosenkai@dlmu.edu.cn}
\par\end{center}

\begin{center}
\emph{Department of Physics, School of Science,}\\
\emph{Dalian Maritime University, }\\
\emph{Dalian 116026, China}\\
\par\end{center}

\vspace{10mm}

\begin{abstract}
In this work, we first derive the corrections to the instanton configuration
of the flavored gauge field in the D4-D8 model with generic flavor
numbers. Then, as the instanton configuration on the D8-branes represents
equivalently baryon in this model, keeping our corrections in hand,
we systemically study the spectrum of baryon, heavy-light baryon or
heavy-light meson and find it is possible to fit the experimental
data with the meson data in this model. Besides, we briefly outline
how to include the interaction of glueball and heavy-light meson or
baryon with our corrections, evaluate numerically the decay rate of
the heavy-light meson or baryonic matter involving glueball. Since
it is possible to fit all the spectra with same choice of the parameters
to the experimental data, we believe our corrections improve the framework
of D4-D8 model and the corrected instanton configuration is also useful
to investigate other properties of baryon in holography.
\end{abstract}
\newpage{}

\tableofcontents{}

\section{Introduction}

QCD as the characteristic strong coupling gauge theory has been expecting
to be analyzed by the gauge-gravity duality and AdS/CFT since holography
of gravity is proposed in 1990s \cite{key-1,key-2,key-3,key-4}. Along
this direction, there are several holographic frameworks for QCD attracting
great interests in the last two decades e.g. \cite{key-5,key-6,key-7,key-8}.
Among these works, the D4-D8 model i.e. the Witten-Sakai-Sugimoto
model \cite{key-9,key-10}, as a top-down approach based on the underlying
string theory, becomes famous and successful in the process of time
because this model includes mostly all the fundamental elements of
QCD in a very simple way so that it could reproduce various elementary
features of QCD e.g. deconfinement and chiral phase transition \cite{key-11,key-12,key-13,key-14,key-15,key-16,key-17,key-18},
baryon spectrum \cite{key-19,key-20,key-21,key-22,key-23,key-24},
glueball spectrum \cite{key-25,key-26,key-27,key-28,key-29}, the
interaction of glueball and meson or baryon \cite{key-30,key-31},
the theta term \cite{key-32,key-33,key-34,key-35,key-36,key-37,key-38}.

Specifically, the D4-D8 model consists of $N_{c}$ coincident D4-branes
as colors and a stack of $N_{f}$ pairs of probe D8- and anti D8-branes
($\mathrm{D8}/\overline{\mathrm{D8}}$-branes) as flavors vertical
to the D4-branes. The open string on the $N_{c}$ D4-branes and $N_{f}$
$\mathrm{D8}/\overline{\mathrm{D8}}$-branes is respectively in the
adjoint representation of $U\left(N_{c}\right)$ and $U\left(N_{f}\right)$
which is therefore identified as gluon and meson. The open string
connecting $N_{c}$ D4-branes and $N_{f}$ $\mathrm{D8}/\overline{\mathrm{D8}}$-branes
is in the fundamental representation of $U\left(N_{c}\right)$ and
$U\left(N_{f}\right)$ which is accordingly identified as the fundamental
chiral quark. In the large $N_{c}$ limit, the bulk geometry is described
by the type IIA supergravity which can be solved by the bubble configuration
of the D4-branes compactified on a circle since the dual theory will
exhibit confinement in this geometry. And the supersymmetry would
break down in the low-energy theory when the periodic and anti-periodic
condition is respectively imposed to the gauge boson and supersymmetric
fermion along the circle, as it is illustrated in Figure \ref{fig:1}.
Follow the idea in Witten's \cite{key-39}, it is possible to introduce
baryon vertex into the D4-D8 model which is a D4-brane wrapped on
$S^{4}$. Analyzing the charge of D4- and D8-brane, the baryon vertex
is recognized as the instanton solution of the gauge field on the
D8-brane \cite{key-40}, hence the Hamiltonian of the collective modes,
whose eigen value would be interpreted as the baryon spectrum, can
be derived by additionally employing the idea of Skyrmions in the
moduli space of instanton \cite{key-41}, as it is discussed in \cite{key-19}.
Beside, the glueball field in this model is identified as the bulk
gravitational polarization since it is sourced by gauge invariant
operator as the energy-momentum tensor of the gluon \cite{key-25,key-26,key-27,key-28,key-29}.
In this sense, when the bulk gravitational polarization is taken into
account, there must be that the bulk close string interacts with the
open string on the D8-branes and open string on the baryon vertex
since the bulk gravitational polarization is excited by the bulk closed
string. Thus the effective action of the D8-brane will include the
coupling term of the bulk gravitational polarization and the gauge
field or instanton, which can be interpreted as the interaction of
glueball and meson or baryon. So when we derive the Hamiltonian of
the collective modes, it will arise a time-dependent term describing
the decay of baryon involving glueball as \cite{key-30,key-31}. Altogether,
the D4-D8 model could be treated basically as a holographic version
of QCD.

Since the subject of this work is the corrections to baryon in the
D4-D8 model, our concern would be the instanton solution of the gauge
field on the D8-brane and the associated Hamiltonian of the collective
modes. The motivation of this work comes from \cite{key-19,key-22,key-23,key-24,key-30,key-31}.
In \cite{key-19}, it turns out the effective action of the D8-brane
in the strong coupling limit (i.e. the 't Hooft coupling $\lambda$
goes to infinity $\lambda\rightarrow\infty$) is pure Yang-Mills action,
so that the instanton solution to the non-Abelian spatial part of
the gauge field can be chosen as the $SU\left(2\right)$ Belavin--Polyakov--Schwarz--Tyupkin
(BPST) solution which represents the Euclidean instanton. However
the baryon spectrum based on the BPST instanton in \cite{key-19}
is unable to fit the experimental data of baryon when the meson data
in this model is employed even if the framework in \cite{key-19}
is generalized into three-flavor case \cite{key-20}. The same problem
also appears in \cite{key-22,key-23} where the meson data in the
D4-D8 model is abandoned. The most likely reason could be that the
derivation in \cite{key-19,key-20,key-22,key-23} is strictly valid
in the limit of $\lambda\rightarrow\infty$ while $\lambda$ is a
finite number in realistic QCD. To figure out this issue, \cite{key-24}
proposes a possible correction to the two-flavored instanton solution
in this model. Since the two-flavored baryon spectrum is unrealistic,
the baryon spectrum with the correction proposed in \cite{key-24}
may not fit the experimental data well enough when the meson data
in this model is picked up. Nonetheless, in this work, we attempt
to generalize the corrections to $SU\left(2\right)$ instanton into
the case of $SU\left(N_{f}\right)$ instanton as \cite{key-20}. Afterwards,
we obtain the corrected baryon spectrum with generic flavor number
$N_{f}$. Take into account the symmetries of isospin and angular
momentum, we find the corrected baryon spectrum with $N_{f}=3$ fits
very well to the experimental data by picking up the meson data in
this model. Moreover, when the heavy flavor is introduced into this
model as \cite{key-22,key-23,key-35}, the corrected heavy-light baryon
spectrum also fits well to the experimental data with the meson data
in this model. And we also derive the $N_{f}=2$ heavy-light baryon
spectrum in order to fit the experimental data of heavy-light meson,
since the heavy-light meson could be treated as quasi-baryon \cite{key-42,key-43,key-44}.
Picking up our corrections, the heavy-light meson spectrum with meson
data in this model fits well to the experimental data of the lowest
D-mesons. Finally, the bulk gravitational polarization is introduced
with our corrections to the instanton, so it is able to describe the
decay of the heavy-light meson involving the glueball in this framework
\footnote{Since the lightest glueball mass is evaluated around 1000MeV by lattice
QCD, it is mostly produced in the decay of baryon or baryonic meson
e.g. \cite{key-45,key-46,key-47}. And it is also a motivation to
include the interaction of glueball and baryon or heavy-light meson
in the framework of D4-D8 model.}. Despite the corrections to the decay rate of the heavy-light baryon
\cite{key-30,key-31}, it is out of reach to fit the experimental
data exactly since the experimental data of glueball is less clear.
Overall, we believe our corrections to the instanton as baryon improve
the framework of D4-D8 approach since it is able to fit the both the
spectra of meson and baryon with same parameters, especially when
the heavy flavor is included.

The outline of this paper is as follows. In Section 2, we collect
the essential parts of the D4-D8 model. In Section 3, we derive our
corrections to the BPST instanton solution with generic $N_{f}$ as
a generalization of \cite{key-24}. Then compare our corrected baryon
spectrum with the experimental data in the case of $N_{f}=3$. In
Section 4, we include the heavy flavor in the D4-D8 model, obtain
the heavy-light baryon spectrum with our corrections and fit the experimental
data of the heavy-light meson. In Section 5, we briefly outline how
to describe the interaction of glueball and baryon or baryonic matter
with our corrections, then evaluate the decay rate of heavy-light
meson involving glueball numerically. The final section is the summary
and discussion.

\section{Baryon as instanton in D4-D8 model}

In this section, let us collect the essential substance of the instanton
configuration and the derivation of baryon spectrum in the D4-D8 model
from \cite{key-9,key-10,key-19,key-20}.

\subsection{The D4-D8 model}

\begin{figure}
\begin{centering}
\includegraphics[scale=0.2]{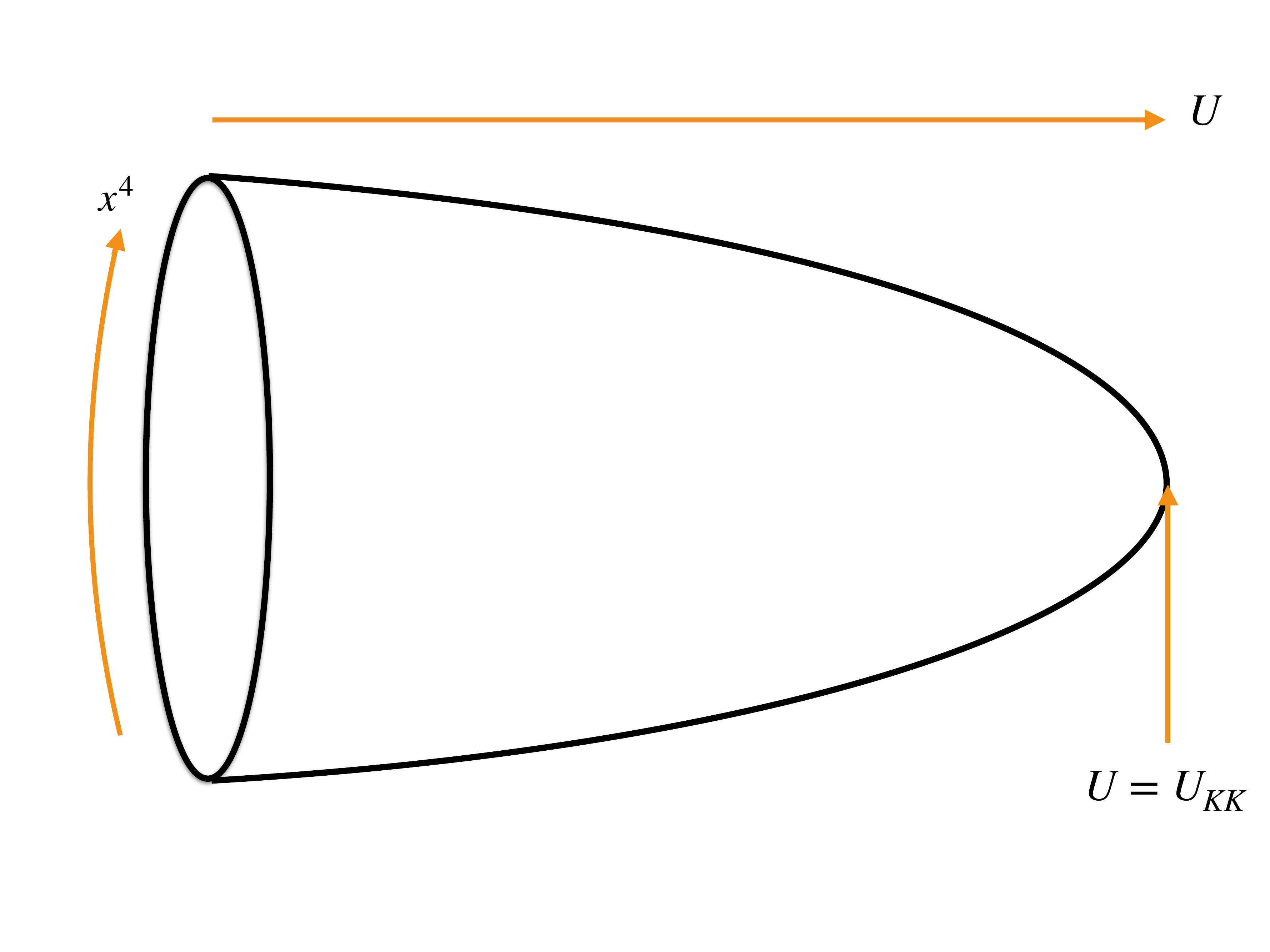}\includegraphics[scale=0.2]{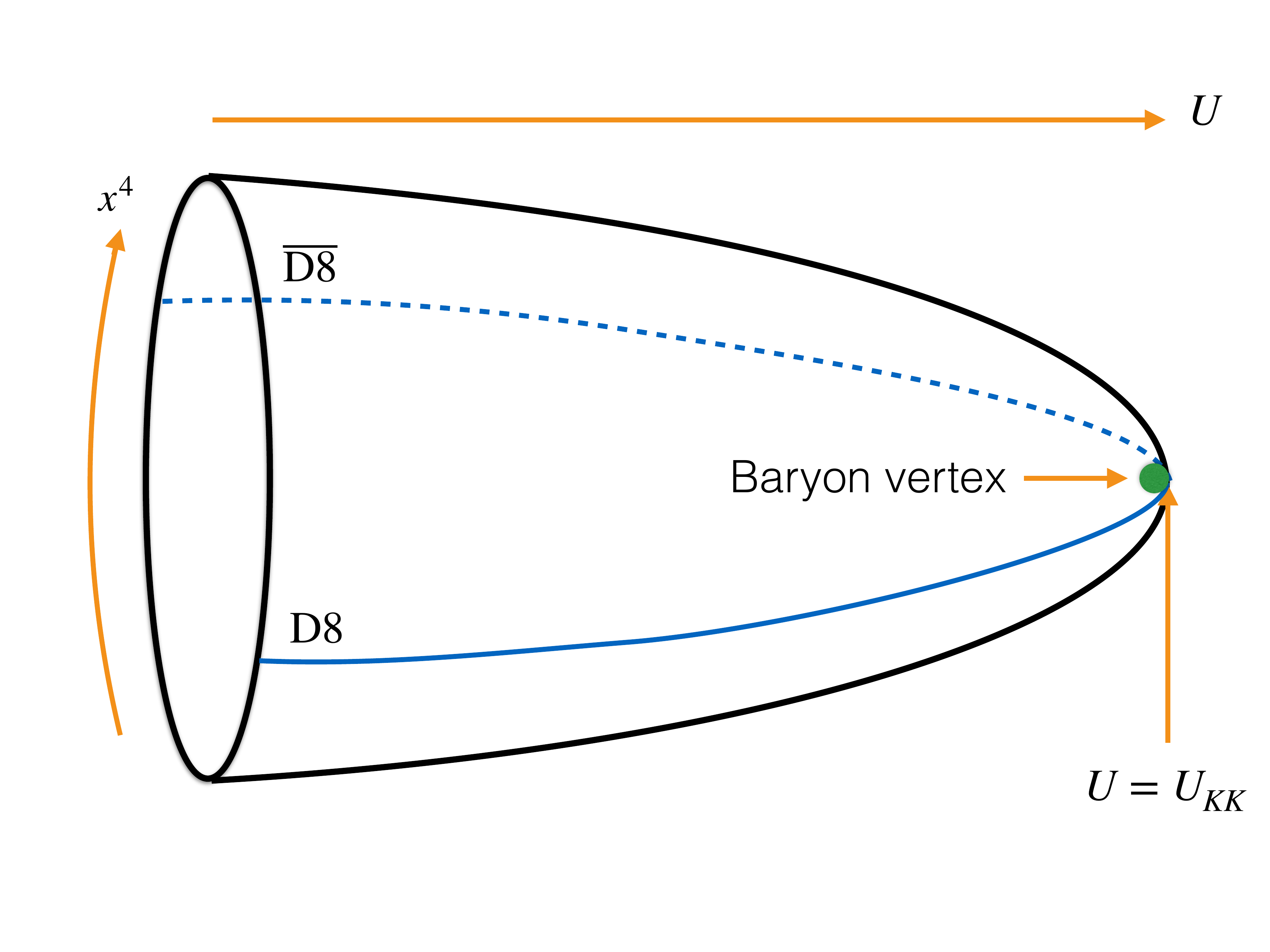}
\par\end{centering}
\caption{\label{fig:1} The D-brane configurations in the D4-D8 model. \textbf{Left:}
The bubble bulk geometry on $\left\{ U,x^{4}\right\} $ plane produced
by $N_{c}$ coincident D4-branes. \textbf{Right:} The $N_{f}$ pairs
of $\mathrm{D8}/\overline{\mathrm{D8}}$-branes (blue line) and baryon
vertex (green point) as probes in the bulk.}
\end{figure}

The D4-D8 model consists of $N_{c}$ coincident D4-branes and a stack
of $N_{f}$ pairs of probe $\mathrm{D8}/\overline{\mathrm{D8}}$-branes
in the large $N_{c}$ limit. The bulk geometry is described by type
IIA supergravity in 10-dimension (10d) given as \cite{key-9},

\begin{align}
ds^{2} & =\left(\frac{U}{R}\right)^{3/2}\left[\eta_{\mu\nu}dx^{\mu}dx^{\nu}+f\left(U\right)\left(dx^{4}\right)^{2}\right]+\left(\frac{R}{U}\right)^{3/2}\left[\frac{dU^{2}}{f\left(U\right)}+U^{2}d\Omega_{4}^{2}\right],\nonumber \\
e^{\phi} & =g_{s}\left(\frac{U}{R}\right)^{3/4},F_{4}=dC_{3}=\frac{2\pi N_{c}}{\Omega_{4}}\epsilon_{4},f\left(U\right)=1-\frac{U_{KK}^{3}}{U^{3}}.\label{eq:1}
\end{align}
This gravity solution describes the bubble configuration of the spacetime
ending on $U=U_{KK}$ as it is illustrated in Figure \ref{fig:1}.
 The D4-branes extend along $\left\{ x^{\mu},x^{4}\right\} $ where
the index $\mu,\nu$ runs over 0,1,2,3. The field $\phi,C_{3}$ is
respectively the dilaton and Ramond-Ramond 3-form in the type IIA
superstring theory. Here $\epsilon_{4}$, $\Omega_{4}=8\pi^{2}/3$
is the volume form, the volume of a unit $S^{4}$ and $R$ refers
to the radius of the bulk which relates to the string coupling $g_{s}$
and string length $l_{s}$ as $R^{3}=\pi g_{s}N_{c}l_{s}^{3}$. Note
that the direction $x^{4}$ is compactified on $S^{1}$ with a period
$\delta x^{4}$ as $x^{4}\sim x^{4}+\delta x^{4}$, so, above the
size $\delta x^{4}$, the supersymmetry is broken down in the low-energy
effective theory on the D4-branes once the periodic and anti-periodic
condition is imposed to the boson and fermion along $S^{1}$ \cite{key-4}.
In order to avoid the conical singularity at $U=U_{KK}$, we can define
the Kaluza-Klein mass $M_{KK}$ as,

\begin{equation}
M_{KK}=\frac{2\pi}{\delta x^{4}}=\frac{3U_{KK}^{1/2}}{2R^{3/2}},
\end{equation}
which specifies the dual theory is effectively four-dimensional confining
Yang-Mills (YM) theory. By examining the dual theory on a probe D4-brane,
the variables in terms of field theory can be expressed as,

\begin{equation}
R^{3}=\frac{1}{2}\frac{g_{YM}^{2}N_{c}l_{s}^{2}}{M_{KK}},\ U_{KK}=\frac{2}{9}g_{YM}^{2}N_{c}M_{KK}l_{s}^{2},\ g_{s}=\frac{1}{2\pi}\frac{g_{YM}^{2}}{M_{KK}l_{s}},
\end{equation}
where $g_{YM}$ refers to the Yang-Mills coupling constant in the
dual theory. 

The $N_{f}$ pairs of probe $\mathrm{D8}/\overline{\mathrm{D8}}$-branes
embedded into the bulk geometry (\ref{eq:1}) are perpendicular and
antipodal to the compactified direction $x^{4}$ as it is displayed
in Figure \ref{fig:1}. The action of D8-brane is given by the Dirac-Born-Infeld
(DBI) plus Wess-Zumino (WZ) action as ,

\begin{align}
S_{\mathrm{D8}} & =S_{\mathrm{DBI}}+S_{\mathrm{WZ}},\nonumber \\
S_{\mathrm{DBI}} & =-T_{8}\int_{\mathrm{D8}/\overline{\mathrm{D8}}}d^{9}xe^{-\phi}\mathrm{STr}\sqrt{-\det\left(g_{\alpha\beta}+2\pi\alpha^{\prime}\mathcal{F}_{\alpha\beta}\right)}\nonumber \\
S_{\mathrm{WZ}} & =\left(2\pi\alpha^{\prime}\right)^{3}T_{8}\int_{\mathrm{D8}/\overline{\mathrm{D8}}}C_{3}\mathrm{Tr}\mathcal{F}^{3},\label{eq:4}
\end{align}
where the index $\alpha,\beta$ runs over the D8-brane, $T_{8}=\left(2\pi\right)^{-8}l_{s}^{-9}$
is the tension of the D8-brane and $\mathcal{F}$ refers to the $U\left(N_{f}\right)$
Yang-Mills gauge field strength on the D8-branes. Expand the DBI action
up to quadratic term and integrate the WZ action by part, the action
(\ref{eq:4}) becomes,

\begin{align}
S_{\mathrm{D8}} & =S_{\mathrm{YM}}\left[\mathcal{A}\right]+S_{\mathrm{CS}}\left[\mathcal{A}\right],\nonumber \\
S_{\mathrm{YM}}\left[\mathcal{A}\right] & =-\kappa\int d^{4}xdz\mathrm{Tr}\left[\frac{1}{2}h\left(z\right)\mathcal{F}_{\mu\nu}^{2}+k\left(z\right)\mathcal{F}_{\mu z}^{2}\right],\nonumber \\
S_{\mathrm{CS}}\left[\mathcal{A}\right] & =\frac{N_{c}}{24\pi^{2}}\int\omega_{5}^{U\left(N_{f}\right)}\left(\mathcal{A}\right),\nonumber \\
h\left(z\right) & =\left(1+z^{2}\right)^{-1/3},k\left(z\right)=1+z^{2}\label{eq:5}
\end{align}
where the formulas are expressed in $M_{KK}=1$ and the parameter
$\kappa$ is given as,

\begin{equation}
\kappa=a\lambda N_{c},\ a=\frac{1}{216\pi^{3}},\lambda=g_{YM}^{2}N_{c}.
\end{equation}
We have used the dimensionless Cartesian coordinate $z$ given by

\begin{equation}
U=U_{KK}\left(1+z^{2}\right)^{1/3}.\label{eq:7}
\end{equation}
Here $\mathcal{A}$ refers to the $U\left(N_{f}\right)$ Yang-Mills
gauge potential associated to $\mathcal{F}$ as $\mathcal{F}=d\mathcal{A}+i\mathcal{A}\wedge\mathcal{A}$
which does not have components along $S^{4}$ and is independent on
$S^{4}$. The gauge Chern-Simons (CS) 5-form $\omega_{5}^{U\left(N_{f}\right)}\left(\mathcal{A}\right)$
is given as,

\begin{equation}
\omega_{5}^{U\left(N_{f}\right)}\left(\mathcal{A}\right)=\mathrm{Tr}\left(\mathcal{A}\mathcal{F}^{2}-\frac{i}{2}\mathcal{A}^{3}\mathcal{F}-\frac{1}{10}\mathcal{A}^{5}\right).\label{eq:8}
\end{equation}
Therefore the following concern is to describe the baryon with the
action presented in (\ref{eq:5}).

\subsection{The classical instanton solution}

According to the gauge-gravity duality, baryon in the D4-D8 model
is recognized as the D4-brane wrapped on $S^{4}$ \cite{key-39} presented
in (\ref{eq:1}) which is illustrated in Figure \ref{fig:1}. On the
other hand, by analyzing the charge of the D4- and D8-brane, in the
D4-D8 model, baryon can be identified as the instanton configuration
of the gauge field on the D8-branes \cite{key-40}. Hence the instanton
solution for the gauge field on D8-brane is the key to describe baryon
in this model.

In order to obtain a low-energy solution representing a baryon for
the gauge field on D8-branes, let us follow the steps in \cite{key-19}.
Specifically, we need an instanton solution for the D8-brane action
(\ref{eq:5}) in the $1/\lambda$ expansion since the \textquoteright t
Hooft coupling $\lambda$ is expected to be large in the dual theory.
To carry out a systematic $1/\lambda$ expansion, we can rescale the
coordinate $\left\{ x^{0},x^{i},z\right\} $ and the gauge field $\mathcal{A}$
as,

\begin{align}
x^{M}\rightarrow\lambda^{-1/2}x^{M},\  & x^{0}\rightarrow x^{0}\nonumber \\
\mathcal{A}_{M}\rightarrow\lambda^{1/2}\mathcal{A}_{M},\  & \mathcal{A}_{0}\rightarrow\mathcal{A}_{0}\nonumber \\
\mathcal{F}_{MN}\rightarrow\lambda\mathcal{F}_{MN},\  & \mathcal{F}_{0M}\rightarrow\lambda^{1/2}\mathcal{F}_{0M},\label{eq:9}
\end{align}
where the indices denoted by capital letters $M,N$ run over 1,2,3,$z$.
Thus the Yang-Mills action in (\ref{eq:5}) can be written as,

\begin{align}
S_{\mathrm{YM}}= & -aN_{c}\int d^{4}xdz\mathrm{Tr}\left[\frac{\lambda}{2}F_{MN}^{2}+\left(-\frac{z^{2}}{6}F_{ij}^{2}+z^{2}F_{iz}^{2}-F_{0M}^{2}\right)+\mathcal{O}\left(\lambda^{-1}\right)\right]\nonumber \\
 & -\frac{aN_{c}}{2}\int d^{4}xdz\left[\frac{\lambda}{2}\hat{F}_{MN}^{2}+\left(-\frac{z^{2}}{6}\hat{F}_{ij}^{2}+z^{2}\hat{F}_{iz}^{2}-\hat{F}_{0M}^{2}\right)+\mathcal{O}\left(\lambda^{-1}\right)\right],\label{eq:10}
\end{align}
while the Chern-Simons action in (\ref{eq:5}) is invariant under
this rescaling. Note that we have decomposed the $U\left(N_{f}\right)$
group as $U\left(N_{f}\right)\simeq U\left(1\right)\times SU\left(N_{f}\right)$
and correspondingly, the generator $\mathcal{A}$ of $U\left(N_{f}\right)$
is decomposed as, 
\begin{equation}
\mathcal{A}=A+\frac{1}{\sqrt{2N_{f}}}\hat{A}=A^{a}t^{a}+\frac{1}{\sqrt{2N_{f}}}\hat{A},\label{eq:11}
\end{equation}
where $\hat{A},A$ refers respectively to the generator of $U\left(1\right)$,
$SU\left(N_{f}\right)$ and $t^{a}$ ($a=1,2...N_{f}^{2}-1$) are
the normalized Hermitian bases of the $su\left(N_{f}\right)$ algebra
satisfying

\begin{equation}
\mathrm{Tr}\left(t^{a}t^{b}\right)=\frac{1}{2}\delta^{ab}.
\end{equation}
In this convention, the Chern-Simons term in (\ref{eq:5}) can be
derived as,

\begin{align}
S_{\mathrm{CS}}= & \frac{N_{c}}{24\pi^{2}}\int\omega_{5}^{SU\left(N_{f}\right)}\left(A\right)+\frac{N_{c}}{24\pi^{2}}\sqrt{\frac{2}{N_{f}}}\epsilon_{MNPQ}\int d^{4}xdz\bigg[\frac{3}{8}\hat{A}_{0}\mathrm{Tr}\left(F_{MN}F_{PQ}\right)\nonumber \\
 & -\frac{3}{2}\hat{A}_{M}\mathrm{Tr}\left(\partial_{0}A_{N}F_{PQ}\right)+\frac{3}{4}\hat{F}_{MN}\mathrm{Tr}\left(A_{0}F_{PQ}\right)+\frac{1}{16}\hat{A}_{0}\hat{F}_{MN}\hat{F}_{PQ}-\frac{1}{4}\hat{A}_{M}\hat{F}_{0N}\hat{F}_{PQ}\nonumber \\
 & +\left(\mathrm{total\ derivatives}\right)\bigg].\label{eq:13}
\end{align}
Then the equations of motion can be obtained by varying the action
(\ref{eq:10}) plus (\ref{eq:13}). For generic $N_{f}\geq2$, the
instanton solution can be obtained by employing the classical $SU\left(2\right)$
BPST solution as embeddable package \cite{key-20} which is given
as,

\begin{equation}
A_{M}^{\mathrm{cl}}=-if\left(\xi\right)g\left(x\right)\partial_{M}g^{-1},\label{eq:14}
\end{equation}
where

\begin{align}
f\left(\xi\right) & =\frac{\xi^{2}}{\xi^{2}+\rho^{2}},\xi=\sqrt{\left(x^{M}-X^{M}\right)^{2}},\nonumber \\
g\left(x\right) & =\left(\begin{array}{cc}
g^{SU\left(2\right)}\left(x\right) & 0\\
0 & \mathbf{1}_{N_{f}-2}
\end{array}\right),g^{SU\left(2\right)}\left(x\right)=\frac{1}{\xi}\left[\left(z-Z\right)\mathbf{1}_{2}-i\left(x^{i}-X^{i}\right)\tau^{i}\right].\label{eq:15}
\end{align}
We use $\mathbf{1}_{N}$ to denote the $N\times N$ identity matrix,
and $\tau^{i}$'s are the Pauli matrices. The constants $X^{M}=\left\{ X^{i},Z\right\} $
and $\rho$ refer respectively to the position and the size of the
instanton which have already been rescaled as (\ref{eq:9}), hence
the $U\left(1\right)$ part of the gauge field can be solved as,

\begin{equation}
\hat{A}_{0}^{\mathrm{cl}}=\sqrt{\frac{2}{N_{f}}}\frac{1}{8\pi^{2}a}\frac{1}{\xi^{2}}\left[1-\frac{\rho^{4}}{\left(\xi^{2}+\rho^{2}\right)^{2}}\right],\ \hat{A}_{M}^{\mathrm{cl}}=0.\label{eq:16}
\end{equation}
which leads to a non-trivial $A_{0}$ as,

\begin{equation}
A_{0}^{\mathrm{cl}}=\frac{1}{16\pi^{2}a}\frac{1}{\xi^{2}}\left[1-\frac{\rho^{4}}{\left(\xi^{2}+\rho^{2}\right)^{2}}\right]\left(\mathcal{P}_{2}-\frac{2}{N_{f}}\mathbf{1}_{N_{f}}\right),
\end{equation}
where $\mathcal{P}_{2}$ is a $N_{f}\times N_{f}$ matrix defined
as $\mathcal{P}_{2}=\mathrm{diag}\left(1,1,0,...0\right)$.

\subsection{Lagrangian of the collective modes and baryon spectrum}

In order to obtain the baryon spectrum, we need to derive the Lagrangian
$L$ of the collective coordinates $\mathcal{X}^{\alpha}$ in the
moduli space of the one-instanton solution. For generic $N_{f}$,
the collective coordinates $\mathcal{X}^{\alpha}$ consist of $\left\{ X^{M},\rho,y^{a}\right\} $,
where $W=y^{a}t^{a}$ is the $SU\left(N_{f}\right)$ orientation of
the instanton. The basic idea here is to approximate the classical
soliton by slowly moving so that the collective coordinates $\mathcal{X}^{\alpha}$
are promoted to become time-dependent as $\mathcal{X}^{\alpha}\left(t\right)$
\cite{key-41}. Thus the Lagrangian of the collective coordinates
is expected to be the element of the world line with a potential in
the moduli space as,

\begin{equation}
L\left(\mathcal{X}^{\alpha}\right)=\frac{m_{X}}{2}\mathcal{G}_{\alpha\beta}\dot{\mathcal{X}}^{\alpha}\dot{\mathcal{X}}^{\beta}-U\left(\mathcal{X}^{\alpha}\right)+\mathcal{O}\left(\lambda^{-1}\right),\label{eq:18}
\end{equation}
where $\mathcal{G}_{\alpha\beta}$ refers to the metric of the moduli
space and the potential $U\left(\mathcal{X}^{\alpha}\right)$ is the
classical soliton mass given by $S\left[\mathcal{A}^{\mathrm{cl}}\right]=-\int dtU\left(\mathcal{X}^{\alpha}\right)$.
By the approximation, the $SU\left(N_{f}\right)$ gauge field is also
expected to be time-dependent by a gauge transformation,

\begin{align}
A_{M}\left(t,x\right) & =W\left(t\right)A_{M}^{\mathrm{cl}}\left(x,\mathcal{X}^{\alpha}\right)W\left(t\right)^{-1}-iW\left(t\right)\partial_{M}W\left(t\right)^{-1},\nonumber \\
A_{0}\left(t,x\right) & =W\left(t\right)A_{0}^{\mathrm{cl}}\left(x,\mathcal{X}^{\alpha}\right)W\left(t\right)^{-1}+\Delta A_{0},\nonumber \\
\hat{A}_{M}\left(t,x\right) & =0,\ \hat{A}_{0}\left(t,x\right)=\hat{A}_{0}^{\mathrm{cl}}\left(t,x\right),
\end{align}
where ``cl'' refers to the BPST instanton solution presented in
Section 2.2 with time-dependent $\mathcal{X}^{\alpha}\left(t\right)$
and the associated field strength becomes,

\begin{align}
F_{MN}= & W\left(t\right)F_{MN}^{\mathrm{cl}}W\left(t\right)^{-1},\nonumber \\
F_{0M}= & W\left(t\right)\left(\dot{\mathcal{X}}^{\alpha}\frac{\partial}{\partial\mathcal{X}^{\alpha}}A_{M}^{\mathrm{cl}}-D_{M}^{\mathrm{cl}}\Sigma-D_{M}^{\mathrm{cl}}A_{0}^{\mathrm{cl}}\right)W\left(t\right)^{-1},\nonumber \\
\hat{F}_{0M}= & \hat{F}_{0M}^{\mathrm{cl}},\ \hat{F}_{MN}=\hat{F}_{MN}^{\mathrm{cl}},
\end{align}
where 

\begin{align}
D_{M}^{\mathrm{cl}}A_{0} & =\partial_{M}A_{0}+i\left[A_{M}^{\mathrm{cl}},A_{0}\right],\nonumber \\
\Sigma & =W\left(t\right)^{-1}\Delta A_{0}W\left(t\right)-i\dot{W}\left(t\right)^{-1}W\left(t\right).
\end{align}
Note that $\Delta A_{0}$ must be determined by its equation of motion
from (\ref{eq:10}) and (\ref{eq:13}) which is,

\begin{equation}
D_{M}^{\mathrm{cl}}\left(\dot{X}^{N}\frac{\partial}{\partial X^{N}}A_{M}^{\mathrm{cl}}+\dot{\rho}\frac{\partial}{\partial\rho}A_{M}^{\mathrm{cl}}-D_{M}^{\mathrm{cl}}\Sigma\right)=0.
\end{equation}
The exact solution for $\Sigma$ can be found in \cite{key-19,key-20}.
Then the Lagrangian of the collective modes is given by

\begin{align}
S\left[\mathcal{A}\right]-S\left[\mathcal{A}^{\mathrm{cl}}\right] & =\int dt\left[L_{\mathrm{YM}}\left(\mathcal{X}^{\alpha}\right)+L_{\mathrm{CS}}\left(\mathcal{X}^{\alpha}\right)\right]=\int dtL\left(\mathcal{X}^{\alpha}\right)\nonumber \\
S_{\mathrm{YM}}\left[\mathcal{A}\right]-S_{\mathrm{YM}}\left[\mathcal{A}^{\mathrm{cl}}\right] & =\int dtL_{\mathrm{YM}}\left(\mathcal{X}^{\alpha}\right),\nonumber \\
S_{\mathrm{CS}}\left[\mathcal{A}\right]-S_{\mathrm{CS}}\left[\mathcal{A}^{\mathrm{cl}}\right] & =\int dtL_{\mathrm{CS}}\left(\mathcal{X}^{\alpha}\right).\label{eq:23}
\end{align}
Therefore we can obtain,

\begin{align}
L\left(\mathcal{X}^{\alpha}\right) & =-M+aN_{c}\mathrm{Tr}\int d^{3}xdz\left(\dot{X}^{N}F_{MN}^{\mathrm{cl}}+\dot{\rho}\frac{\partial}{\partial\rho}A_{M}-\dot{X}^{N}D_{M}^{\mathrm{cl}}A_{N}^{\mathrm{cl}}-D_{M}^{\mathrm{cl}}\Sigma\right)^{2}+\mathcal{O}\left(\lambda^{-1}\right)\nonumber \\
 & =-M_{0}+\frac{m_{X}}{2}\delta_{ij}\dot{X}^{i}\dot{X}^{j}+L_{Z}+L_{\rho}+L_{\rho W}+\mathcal{O}\left(\lambda^{-1}\right),\label{eq:24}
\end{align}
where

\begin{align}
L_{Z}= & \frac{m_{Z}}{2}\left(\dot{Z}^{2}-\omega_{Z}^{2}Z^{2}\right),L_{\rho}=\frac{m_{\rho}}{2}\left(\dot{\rho}-\omega_{\rho}^{2}\rho^{2}\right)-\frac{K}{m_{\rho}\rho^{2}},\nonumber \\
L_{\rho W}= & \frac{m_{\rho}\rho^{2}}{2}\sum_{a}C_{a}\left[\mathrm{Tr}\left(-iW^{-1}\dot{W}t^{a}\right)\right]^{2},a=1,2...N_{f}^{2}-1
\end{align}
and

\begin{equation}
M_{0}=8\pi^{2}\kappa,m_{X}=m_{Z}=\frac{m_{\rho}}{2}=8\pi^{2}\kappa\lambda^{-1},K=\frac{2}{5}N_{c}^{2},\omega_{Z}^{2}=4\omega_{\rho}^{2}=\frac{2}{3}.
\end{equation}
Note that we have written the formulas in the unit $M_{KK}=1$ and
the metric of the moduli space can be obtained by comparing (\ref{eq:24})
with (\ref{eq:18}). $C_{a}$'s are constants dependent on the $SU\left(N_{f}\right)$
instanton solution. For example, for $N_{f}=2$, $C_{1,2,3}=1$; for
$N_{f}=3$, $C_{1,2,3}=1,C_{4,5,6,7}=1/2$ and $C_{8}=0$. Accordingly,
the collective modes, as baryon states, can be obtained by quantizing
the Lagrangian (\ref{eq:24}), i.e. replace straightforwardly the
derivative term by $\dot{X}^{\alpha}\rightarrow-\frac{i}{m_{X}}\partial_{\alpha}$.
Afterwards, the quantized Hamiltonian associated to (\ref{eq:24})
is collected as,

\begin{align}
H= & M_{0}+H_{Z}+H_{\rho}+H_{\rho W},\nonumber \\
H_{Z}= & -\frac{1}{2m_{Z}}\partial_{Z}^{2}+\frac{1}{2}m_{Z}\omega_{Z}^{2}Z^{2},\nonumber \\
H_{\rho}= & -\frac{1}{2m_{\rho}}\frac{1}{\rho^{\eta}}\partial_{\rho}\left(\rho^{\eta}\partial_{\rho}\right)+\frac{1}{2}m_{\rho}\omega_{\rho}^{2}\rho^{2}+\frac{K}{m_{\rho}\rho^{2}},\nonumber \\
H_{\rho W}= & \frac{m_{\rho}\rho^{2}}{2}\sum_{a}C_{a}\left[\mathrm{Tr}\left(-iW^{-1}\dot{W}t^{a}\right)\right]^{2}=\frac{2}{m_{\rho}\rho^{2}}\sum_{a}C_{a}\left(J^{a}\right)^{2},\label{eq:27}
\end{align}
where $\eta=N_{f}^{2}-1$ and $J^{a}$'s refer to the operators of
the angular momentum of $SU\left(N_{f}\right)$. Therefore, the baryon
spectrum can be obtained by evaluating the eigen values of the Hamiltonian
(\ref{eq:27}).

\section{Corrections of $\mathcal{O}\left(\lambda^{-1/3}\right)$ to the holographic
baryon}

As we have outlined that baryon in the D4-D8 model can be identified
as the BPST instanton configuration on the D8-brane as its classical
description, let us introduce a possible correction to the BPST solution
presented in Section 2 as the deformed description of holographic
baryon, then analyze the corrected baryon spectrum in this section.

\subsection{Corrections to the classical solution}

We start with equations of motion for the $SU\left(N_{f}\right)$
gauge fields $A_{0},A_{M}$ which are obtained by varying action (\ref{eq:10})
plus (\ref{eq:13}), as,

\begin{align}
D_{M}F_{0M}+\frac{1}{64\pi^{2}a}\sqrt{\frac{2}{N_{f}}}\epsilon_{MNPQ}\hat{F}_{MN}F_{PQ}\nonumber \\
+\frac{1}{64\pi^{2}a}\epsilon_{MNPQ}\left[F_{MN}F_{PQ}-\frac{1}{N_{f}}\mathrm{Tr}\left(F_{MN}F_{PQ}\right)\right]+\mathcal{O}\left(\lambda^{-1}\right) & =0,\label{eq:28}\\
D_{N}F_{MN}+\mathcal{O}\left(\lambda^{-1}\right) & =0,\label{eq:29}\\
\partial_{M}\hat{F}_{0M}+\frac{1}{64\pi^{2}a}\sqrt{\frac{2}{N_{f}}}\epsilon_{MNPQ}\left[\mathrm{Tr}\left(F_{MN}F_{PQ}\right)+\frac{1}{2}\hat{F}_{MN}\hat{F}_{PQ}\right]+\mathcal{O}\left(\lambda^{-1}\right) & =0,\label{eq:30}\\
\partial_{N}\hat{F}_{MN}+\mathcal{O}\left(\lambda^{-1}\right) & =0,\label{eq:31}
\end{align}
where the covariant derivative is defined as $D_{M}A_{N}=\partial_{M}A_{N}+i\left[A_{M},A_{N}\right]$
in our convention. Then let us add the correction to the spatial part
of $SU\left(N_{f}\right)$ BPST solution (\ref{eq:14}) first as,

\begin{align}
\tilde{A}_{M} & =A_{M}^{\mathrm{cl}}+\delta A_{M},\tilde{F}_{MN}=F_{MN}^{\mathrm{cl}}+\delta F_{MN},
\end{align}
where

\begin{align}
\delta F_{MN} & =D_{M}\delta A_{N}-D_{N}\delta A_{M}+i\left[\delta A_{M},\delta A_{N}\right].
\end{align}
In this sense, the equation of motion for $\tilde{A}_{M}$ takes the
same formula as they are given in (\ref{eq:29}) after replacing $D_{M},F_{MN}$
by $\tilde{D}_{M},\tilde{F}_{MN}$, so it leads to $\tilde{D}_{N}\tilde{F}_{MN}=0$
or equivalently,

\begin{equation}
D_{N}D_{N}\delta A_{M}-2i\left[F_{NM}^{\mathrm{cl}},\delta A_{N}\right]=0,\label{eq:34}
\end{equation}
where we have imposed

\begin{equation}
D_{N}F_{MN}^{\mathrm{cl}}=0,\ \left[D_{N},D_{M}\right]\delta A_{P}=i\left[F_{NM},\delta A_{P}\right].
\end{equation}
Besides, the instanton solution $A_{M}^{\mathrm{cl}}$ in (\ref{eq:10})
is gauged by $D_{M}A_{M}^{\mathrm{cl}}=0$ which must remain as $\tilde{D}_{M}\tilde{A}_{M}=0$.
Thus we can obtain the gauge condition for $\delta A_{M}$ as,

\begin{equation}
D_{M}\delta A_{M}=0.\label{eq:36}
\end{equation}
Solve the equation (\ref{eq:34}) with (\ref{eq:36}), we can obtain
a solution for $\delta A_{M}$ as,

\begin{equation}
\delta A_{i}=\frac{1}{2}\frac{B}{\left(\xi^{2}+\rho^{2}\right)^{2}}\delta_{ij}t^{j},\ \delta A_{z}=0,\label{eq:37}
\end{equation}
which is an embedding solution of the corrections to the $SU\left(2\right)$
case presented in \cite{key-24}. The constant $B$ must be determined
by minimizing the classical Yang-Mills plus Chern-Simons action given
in (\ref{eq:10}) (\ref{eq:13}). 

Next, we need to solve the $U\left(1\right)$ part equation (\ref{eq:30})
by picking up a correction $\delta\hat{A}_{0}$ and $\delta\hat{A}_{M}$.
Due to $\hat{A}_{M}^{\mathrm{cl}}=0$, we can simply choose 

\begin{equation}
\delta\hat{A}_{M}=0,\label{eq:38}
\end{equation}
which leads to an equation for $\delta\hat{A}_{0}$ as,

\begin{equation}
\partial_{M}^{2}\left(\hat{A}_{0}^{\mathrm{cl}}+\delta\hat{A}_{0}\right)=\frac{1}{64\pi^{2}a}\sqrt{\frac{2}{N_{f}}}\epsilon_{MNPQ}\mathrm{Tr}\left(\tilde{F}_{MN}\tilde{F}_{PQ}\right).\label{eq:39}
\end{equation}
Using (\ref{eq:14}) (\ref{eq:15}) and (\ref{eq:37}), we can solve
(\ref{eq:39}) as,

\begin{align}
\delta\hat{A}_{0}= & -\sqrt{\frac{2}{N_{f}}}\bigg\{\frac{B}{32\pi^{2}a}\frac{\left(z-Z\right)\left(3\rho^{2}+\xi^{2}\right)}{\rho^{2}\left(\rho^{2}+\xi^{2}\right)^{3}}\nonumber \\
 & +\frac{B^{2}}{1536\pi^{2}a}\frac{1}{\rho^{4}\left(\rho^{2}+\xi^{2}\right)^{4}}\left[9\rho^{4}+\xi^{4}-4\left(z-Z\right)^{2}\xi^{2}+4\rho^{2}\xi^{2}-16\left(z-Z\right)^{2}\rho^{2}\right]\nonumber \\
 & -\frac{B^{3}\left(z-Z\right)}{10240\pi^{2}a\rho^{8}\left(\rho^{2}+\xi^{2}\right)^{5}}\left[11\rho^{6}-3\rho^{2}\xi^{4}+\left(\rho^{2}+\xi^{2}\right)^{2}\left(9\rho^{2}+2\xi^{2}\right)\right]\bigg\}.\label{eq:40}
\end{align}
Finally, by imposing (\ref{eq:37}) - (\ref{eq:40}) into (\ref{eq:28}),
it leads to an equation for $\delta A_{0}$ as,

\begin{equation}
\tilde{D}_{M}\tilde{F}_{0M}+\frac{1}{64\pi^{2}a}\epsilon_{MNPQ}\left[\tilde{F}_{MN}\tilde{F}_{PQ}-\frac{1}{N_{f}}\mathrm{Tr}\left(\tilde{F}_{MN}\tilde{F}_{PQ}\right)\right]=0,\label{eq:41}
\end{equation}
where its second part is calculated as,

\begin{align}
 & \frac{1}{64\pi^{2}a}\epsilon_{MNPQ}\left[\tilde{F}_{MN}\tilde{F}_{PQ}-\frac{1}{N_{f}}\mathrm{Tr}\left(\tilde{F}_{MN}\tilde{F}_{PQ}\right)\right]\nonumber \\
= & \bigg[-\frac{3\rho^{4}}{2\pi^{2}a\left(\xi^{2}+\rho^{2}\right)^{4}}+\frac{3\left(z-Z\right)\rho^{2}B}{2\pi^{2}a\left(\xi^{2}+\rho^{2}\right)^{5}}-\frac{10\left(z-Z\right)^{2}+2\xi^{2}-3\rho^{2}}{32\pi a^{2}\left(\xi^{2}+\rho^{2}\right)^{6}}B^{2}\nonumber \\
 & -\frac{3\left(z-Z\right)B^{3}}{64\pi^{2}a\left(\xi^{2}+\rho^{2}\right)^{7}}\bigg]\left(\mathcal{P}_{2}-\frac{2}{N_{f}}\mathbf{1}_{N_{f}}\right).
\end{align}
Notice that

\begin{align}
\tilde{D}_{M}\tilde{F}_{0M} & =-D_{M}D_{M}A_{0}^{\mathrm{cl}}+D_{M}\delta F_{0M}+i\left[\delta A_{M},F_{0M}^{\mathrm{cl}}\right]+i\left[\delta A_{M},\delta F_{0M}\right]\nonumber \\
\delta F_{0M} & =-\partial_{M}\delta A_{0}+i\left[\delta A_{0},A_{M}\right]+i\left[\delta A_{0},\delta A_{M}\right],
\end{align}
so the equation (\ref{eq:41}) can be rewritten as,
\begin{align}
 & D_{M}\delta F_{0M}+i\left[\delta A_{M},F_{0M}^{\mathrm{cl}}\right]+i\left[\delta A_{M},\delta F_{0M}\right]\nonumber \\
= & -\bigg[\frac{3\left(z-Z\right)\rho^{2}B}{2\pi^{2}a\left(\xi^{2}+\rho^{2}\right)^{5}}-\frac{10\left(z-Z\right)^{2}+2\xi^{2}-3\rho^{2}}{32\pi a^{2}\left(\xi^{2}+\rho^{2}\right)^{6}}B^{2}-\frac{3\left(z-Z\right)B^{3}}{64\pi^{2}a\left(\xi^{2}+\rho^{2}\right)^{7}}\bigg]\nonumber \\
 & \times\left(\mathcal{P}_{2}-\frac{2}{N_{f}}\mathbf{1}_{N_{f}}\right).\label{eq:44}
\end{align}
It seems very difficult to search for a solution for (\ref{eq:44})
due to the presence of the commutators, however the instanton solution
presented in Section 2 implies the commutation relationship $\left[A_{0},A_{M}\right]=0$
and it must remain for $\tilde{A}_{0},\tilde{A}_{M}$ as $\left[\tilde{A}_{0},\tilde{A}_{M}\right]=0$.
Therefore, all commutators should vanish in (\ref{eq:44}) which leads
to the following ansatz for $\delta A_{0}$ as,

\begin{equation}
\delta A_{0}=Q\left(x^{M}\right)\left(\mathcal{P}_{2}-\frac{2}{N_{f}}\mathbf{1}_{N_{f}}\right).\label{eq:45}
\end{equation}
In this sense, the equation (\ref{eq:44}) can be solved as,

\begin{align}
\delta A_{0}= & -\bigg\{\frac{B}{64\pi^{2}a}\frac{\left(z-Z\right)\left(3\rho^{2}+\xi^{2}\right)}{\rho^{2}\left(\rho^{2}+\xi^{2}\right)^{3}}\nonumber \\
 & +\frac{B^{2}}{3072\pi^{2}a}\frac{1}{\rho^{4}\left(\rho^{2}+\xi^{2}\right)^{4}}\left[9\rho^{4}+\xi^{4}-4\left(z-Z\right)^{2}\xi^{2}+4\rho^{2}\xi^{2}-16\left(z-Z\right)^{2}\rho^{2}\right]\nonumber \\
 & -\frac{B^{3}\left(z-Z\right)}{20480\pi^{2}a\rho^{8}\left(\rho^{2}+\xi^{2}\right)^{5}}\left[11\rho^{6}-3\rho^{2}\xi^{4}+\left(\rho^{2}+\xi^{2}\right)^{2}\left(9\rho^{2}+2\xi^{2}\right)\right]\bigg\}\nonumber \\
 & \times\left(\mathcal{P}_{2}-\frac{2}{N_{f}}\mathbf{1}_{N_{f}}\right).
\end{align}
Afterwards, the classical mass of the soliton with the corrections
$\delta\mathcal{A}$ can be evaluated by using $S\left[\mathcal{A}^{\mathrm{cl}}+\delta\mathcal{A}\right]=-\int\left(M+\Delta M\right)dt$,
which, after some straightforward buy very messy calculations, is,

\begin{align}
M+\Delta M= & \kappa\int d^{3}xdz\mathrm{Tr}\left[\frac{1}{2}\tilde{F}_{MN}^{2}+\lambda^{-1}\left(-\frac{z^{2}}{6}\tilde{F}_{ij}^{2}+z^{2}\tilde{F}_{iz}^{2}-\tilde{F}_{0M}^{2}\right)-\frac{\lambda^{-1}}{2}\left(\hat{F}_{0M}^{\mathrm{cl}}+\delta\hat{F}_{0M}\right)^{2}\right]\nonumber \\
 & -\frac{\kappa}{24\pi^{2}a}\lambda^{-1}\int d^{3}xdz\epsilon_{MNPQ}\bigg[\sqrt{\frac{2}{N_{f}}}\frac{3}{8}\left(\hat{A}_{0}+\delta\hat{A}_{0}\right)\mathrm{Tr}\left(\tilde{F}_{MN}\tilde{F}_{PQ}\right)\nonumber \\
 & +\frac{3}{4}\mathrm{Tr}\left(\tilde{A}_{0}\tilde{F}_{MN}\tilde{F}_{PQ}\right)\bigg]+\mathcal{O}\left(\lambda^{-1}\right)\nonumber \\
= & 8\pi^{2}\kappa+\frac{\pi^{2}\kappa}{448\rho^{12}}B^{4}+\frac{8\pi^{2}\kappa}{\lambda}\bigg(\frac{\rho^{2}}{6}+\frac{Z^{2}}{3}+\frac{1}{320\pi^{4}a^{2}\rho^{2}}-\frac{BZ}{12\rho^{2}}+\frac{3B^{2}}{640\rho^{4}}+\frac{B^{2}Z^{2}}{80\rho^{6}}\nonumber \\
 & -\frac{B^{2}}{7\times2^{12}a^{2}\pi^{4}\rho^{8}}-\frac{B^{3}}{1920\rho^{8}}-\frac{B^{4}}{35\times3\times2^{10}\rho^{10}}-\frac{B^{4}Z^{2}}{21\times2^{9}\rho^{12}}-\frac{11B^{4}}{7\times3^{3}\times2^{18}}\bigg),\label{eq:47}
\end{align}
and the terms of $\tilde{F}_{0M}^{2},\frac{3}{4}\mathrm{Tr}\left(\tilde{A}_{0}\tilde{F}_{MN}\tilde{F}_{PQ}\right)$
are absent in \cite{key-24}. By minimizing (\ref{eq:47}), the constant
$B$ is obtained as,

\begin{equation}
B=4\times\left(\frac{7}{6}\right)^{1/3}Z^{1/3}\rho^{10/3}\lambda^{-1/3}+\mathcal{O}\left(\lambda^{-2/3}\right).
\end{equation}
Thus $M,\Delta M$ is respectively evaluated as,

\begin{align}
M & =8\pi^{2}\kappa+\frac{8\pi^{2}\kappa}{\lambda}\left(\frac{\rho^{2}}{6}+\frac{Z^{2}}{3}+\frac{1}{320\pi^{4}a^{2}\rho^{2}}\right),\nonumber \\
\Delta M & =-2\pi^{2}\kappa\lambda^{-4/3}\left(\frac{7}{6}\right)^{1/3}\left(\rho Z\right)^{4/3},\label{eq:49}
\end{align}
which however is independent on $N_{f}$ in our holographic setup.

\subsection{Corrections to the baryon spectrum}

In this section, picking up the corrections $\delta\mathcal{A}$ to
the BPST solution, let us correct the baryon spectrum with $N_{f}=3$
for the realistic case \footnote{Our calculation also covers the results in \cite{key-24} where the
corrected baryon spectrum with $N_{f}=2$ can be reviewed.}. As it is outlined in Section 2, the Lagrangian of the collective
modes can be obtained by using (\ref{eq:23}) (\ref{eq:24}), thus
the total quantized hamiltonian $H_{\mathrm{tot}}$ of the collective
modes with the corrections can be obtained by repeating the calculations
in Section 2 while we need to replace $\mathcal{A}^{\mathrm{cl}}\rightarrow\mathcal{A}^{\mathrm{cl}}+\delta\mathcal{A},\mathcal{F}^{\mathrm{cl}}\rightarrow\mathcal{F}^{\mathrm{cl}}+\delta\mathcal{F}$
with $D_{0,M}\rightarrow\tilde{D}_{0,M}$. Resultantly, it leads to\footnote{By imposing our correction, $\omega_{\rho}$ may also contain a correction
of $\mathcal{O}\left(\lambda^{-5/3}\right)$ which has been neglected. } 

\begin{equation}
H_{\mathrm{tot}}=H+\Delta H+\mathcal{O}\left(\lambda^{-2/3}\right),
\end{equation}
where, for $N_{f}=3$,

\begin{align}
H= & M_{0}+H_{Z}+H_{\rho}\nonumber \\
H_{Z}= & -\frac{1}{2m_{Z}}\partial_{Z}^{2}+\frac{1}{2}m_{Z}\omega_{Z}^{2}Z^{2},\nonumber \\
H_{\rho}= & -\frac{1}{2m_{\rho}}\frac{1}{\rho^{8}}\partial_{\rho}\left(\rho^{8}\partial_{\rho}\right)+\frac{1}{2}m_{\rho}\omega_{\rho}^{2}\rho^{2}+\frac{K^{\prime}}{m_{\rho}\rho^{2}},\nonumber \\
\Delta H= & \Delta M=-2\pi^{2}\kappa\lambda^{-4/3}\left(\frac{7}{6}\right)^{1/3}\left(\rho Z\right)^{4/3}.\label{eq:51}
\end{align}
We have taken the $\left(p,q\right)$ representation for the two $SU\left(3\right)_{J}$
and $SU\left(3\right)_{I}$ which refers respectively to the rotation
and flavor (isospin) symmetry in the Hamiltonian as,

\begin{align}
\sum_{a=1}^{8}\left(J_{a}\right)^{2}= & \frac{1}{3}\left[p^{2}+q^{2}+qp+3\left(p+q\right)\right],\nonumber \\
\sum_{a=1}^{8}\left(J_{a}\right)^{2}= & j\left(j+1\right),\nonumber \\
K^{\prime}= & \frac{N_{c}^{2}}{15}+\frac{4}{3}\left[p^{2}+q^{2}+qp+3\left(p+q\right)\right]-2j\left(j+1\right).
\end{align}
Besides, for $N_{f}=3$, we note that the baryon states with right
spins should be selected by the constraint of the hypercharge,

\begin{equation}
J_{8}=\frac{N_{c}}{2\sqrt{3}},\label{eq:53}
\end{equation}
from the Chern-Simons term. However the Chern-Simons term given in
(\ref{eq:5}) is unable to reach this goal since $L_{\mathrm{CS}}$
in (\ref{eq:23}) would be vanished \cite{key-20}. To figure out
this problem, \cite{key-20,key-48} proposed a new Chern-Simons term
as,

\begin{equation}
S_{\mathrm{CS}}^{\mathrm{new}}=S_{\mathrm{CS}}+\frac{1}{10}\int_{N_{5}}\mathrm{Tr}\left(h^{-1}dh\right)^{5}+\int_{\partial M_{5}}\alpha_{4}\left(dh,\mathcal{A}\right),\label{eq:54}
\end{equation}
where $S_{\mathrm{CS}}$ refers to the Chern-Simons term given in
(\ref{eq:5}). And $N_{5}$ denotes a 5-dimensional manifold whose
boundaries satisfies $\partial N_{5}=\partial M_{5}=M_{4,z=+\infty}-M_{4,z=-\infty}$
with the asymptotics of the gauge field on the D8-branes as,

\begin{equation}
\mathcal{A}|_{z\rightarrow\pm\infty}=h^{\pm}\left(d+\mathcal{A}\right)h^{\pm-1},h|_{\partial M_{5}}=\left(h^{+},h^{-}\right),
\end{equation}
where $\mathcal{A}$ is assumed to be regular on $M_{5}$ and produces
no-boundary contributions \footnote{The formula of $\alpha_{4}$ is given in \cite{key-48}.}.
Accordingly, the constraint of the hypercharge (\ref{eq:53}) could
be produced with the new Chern-Simons term (\ref{eq:54}).

Afterwards, the spectrum of the total Hamiltonian $H_{\mathrm{tot}}$
can be obtained approximately by solving the eigen equation of $H$
with a perturbation $\Delta H$ given in (\ref{eq:51}) and the constraint
(\ref{eq:53}). The eigen functions and values of $H_{Z}$ are nothing
but the eigen functions and values of harmonic oscillator while the
eigen functions of $H_{\rho}$ given by $\psi\left(\rho\right)$ could
be solved as,

\begin{equation}
\psi\left(\rho\right)=e^{-v/2}v^{\beta}\gamma\left(v\right),v=m_{\rho}\omega_{\rho}\rho^{2},\beta=\frac{1}{4}\sqrt{\left(\eta-1\right)^{2}+8K^{\prime}}-\frac{1}{4}\left(\eta-1\right),\label{eq:56}
\end{equation}
where $\gamma\left(v\right)$ is hypergeometrical function satisfying
the following hypergeometrical differential equation,

\begin{equation}
\left[v\frac{d}{dv^{2}}+\left(2\beta+\frac{\eta+1}{2}-v\right)\frac{d}{dv}+\left(\frac{E_{\rho}}{2\omega_{\rho}}-\beta-\frac{\eta+1}{4}\right)\right]\gamma\left(v\right)=0.
\end{equation}
So the eigen value $E_{\rho}$ is solved as

\begin{equation}
E_{\rho}=\omega_{\rho}\left[2n_{\rho}+\frac{1}{2}\sqrt{\left(\eta-1\right)^{2}+8K^{\prime}}+1\right].
\end{equation}
Therefore the total spectrum $E$ of $H$ is given by (in the unit
of $M_{KK}$)

\begin{equation}
E=8\pi^{2}\kappa+\omega_{\rho}\left[2n_{\rho}+\frac{1}{2}\sqrt{\left(\eta-1\right)^{2}+8K^{\prime}}+1\right]+\omega_{Z}\left(n_{Z}+\frac{1}{2}\right),\ n_{\rho},n_{Z}=0,1,2,3...\label{eq:59}
\end{equation}
And using the standard method in quantum mechanics, the leading order
correction to the spectrum (\ref{eq:59}) is given by

\begin{equation}
\Delta E=\left\langle \Delta H\right\rangle ,
\end{equation}
which leads to the approximated spectrum of $H_{\mathrm{tot}}$. To
simply compare our corrected baryon spectrum with the realistic QCD,
we could set $N_{c}=3$, so the constraint (\ref{eq:53}) requires
that $\left(p,q\right)$ must satisfy

\begin{equation}
p+2q=3\times\left(\mathrm{integer}\right).
\end{equation}
Therefore the allowed states with smaller $\left(p,q\right)$, $j$
and $K^{\prime}$ are given as,

\begin{align}
\left(p,q\right)=\left(1,1\right), & j=\frac{1}{2},K^{\prime}=\frac{111}{10},\left(\mathrm{octet}\right)\nonumber \\
\left(p,q\right)=\left(3,0\right), & j=\frac{3}{2},K^{\prime}=\frac{171}{10},\left(\mathrm{decuplet}\right)\nonumber \\
\left(p,q\right)=\left(0,3\right), & j=\frac{1}{2},K^{\prime}=\frac{231}{10},\left(\mathrm{\mathrm{anti-decuplet}}\right).
\end{align}
Keeping above in mind, while the unit $M_{KK}$ is not undetermined
in this model, it is possible to compare the experimental data with
our holographic baryon spectrum as $E+\Delta E$. In order to fit
our baryon spectrum to the experimental data, we additionally notice
that, on the other hand, the D4-D8 model is also able to give the
meson spectrum which requests for the parameters $M_{KK}=949\mathrm{MeV},\lambda=16.6$
(the only parameters in our theory) for a realistic matching \cite{key-9}.
Hence let us employ the same choice of $M_{KK},\lambda$, as the meson
data in this model, for the lowest octet and the decuplet or anti-decuplet
baryons $\left(n_{\rho},n_{Z}\right)=\left(0,0\right)$, then the
mass difference is evaluated with our corrections as,

\begin{align}
M_{\mathbf{10}}-M_{\mathbf{8}} & =299.6\mathrm{MeV},\nonumber \\
M_{\mathbf{10}^{*}}-M_{\mathbf{8}} & =564.7\mathrm{MeV},
\end{align}
which is very close to the experimental data

\begin{align}
M_{\mathbf{10}}^{\mathrm{exp}}-M_{\mathbf{8}}^{\mathrm{exp}} & \simeq292\mathrm{MeV},\nonumber \\
M_{\mathbf{10}^{*}}^{\mathrm{exp}}-M_{\mathbf{8}}^{\mathrm{exp}} & \simeq590.7\mathrm{MeV},
\end{align}
from the Particle Data Group (PDG) \cite{key-49} and we have used
the $\Theta^{+}$ mass of $1530\mathrm{MeV}$ as the lowest anti-decuplet
baryon. In this sense, we believe this is a good correction to the
framework of D4-D8 approach since both meson and baryon spectrum could
be fit well.

\section{The heavy flavor}

In this section, we will first outline how to include the heavy flavor
in the D4-D8 model by employing the Higgs mechanism in string theory.
Then let us obtain the heavy-light baryon spectrum with our corrections
to the BPST instanton solution.

\subsection{Higgs mechanism and the massive flavor in the D4-D8 model}

\begin{figure}
\begin{centering}
\includegraphics[scale=0.2]{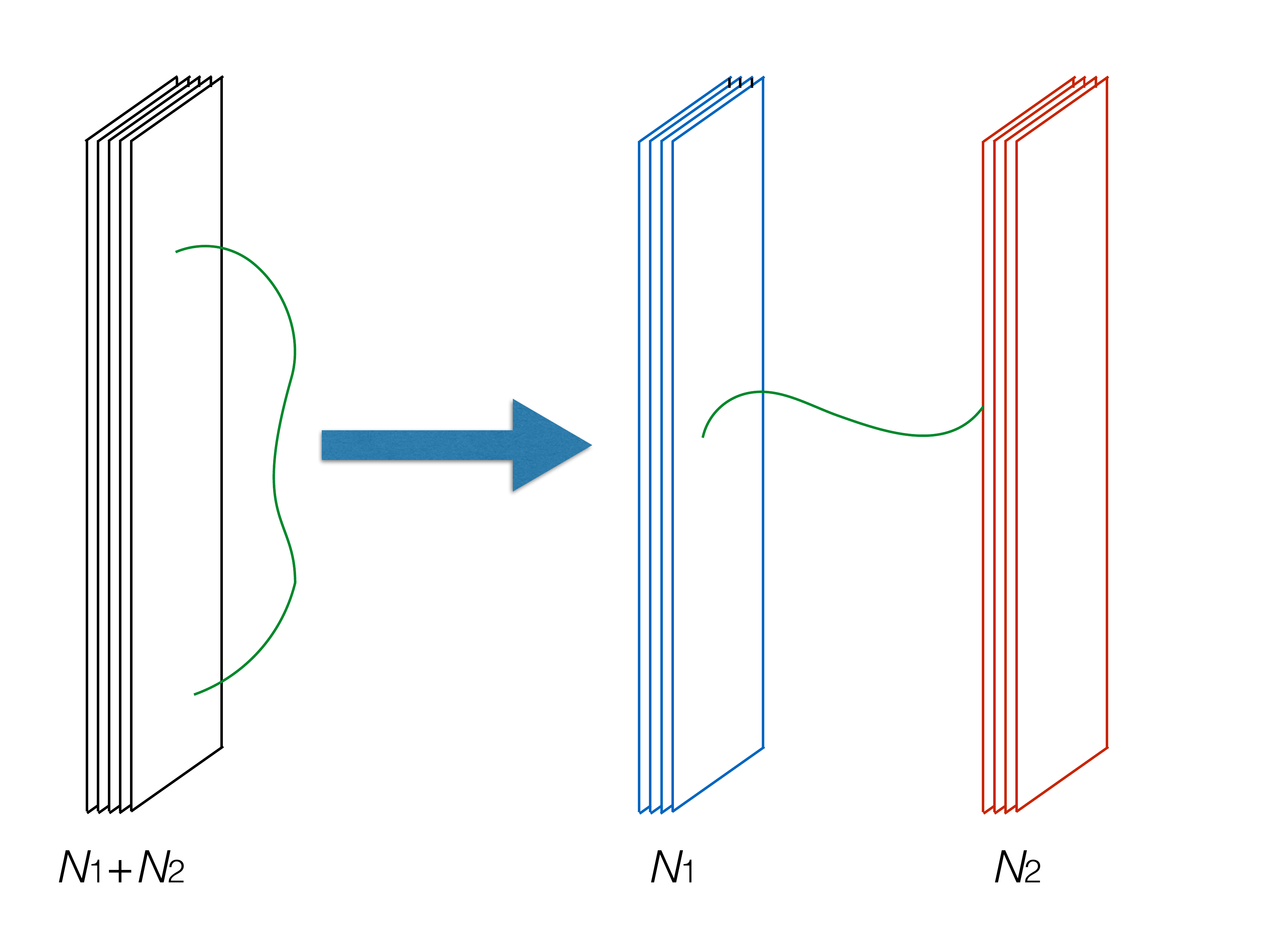}\includegraphics[scale=0.2]{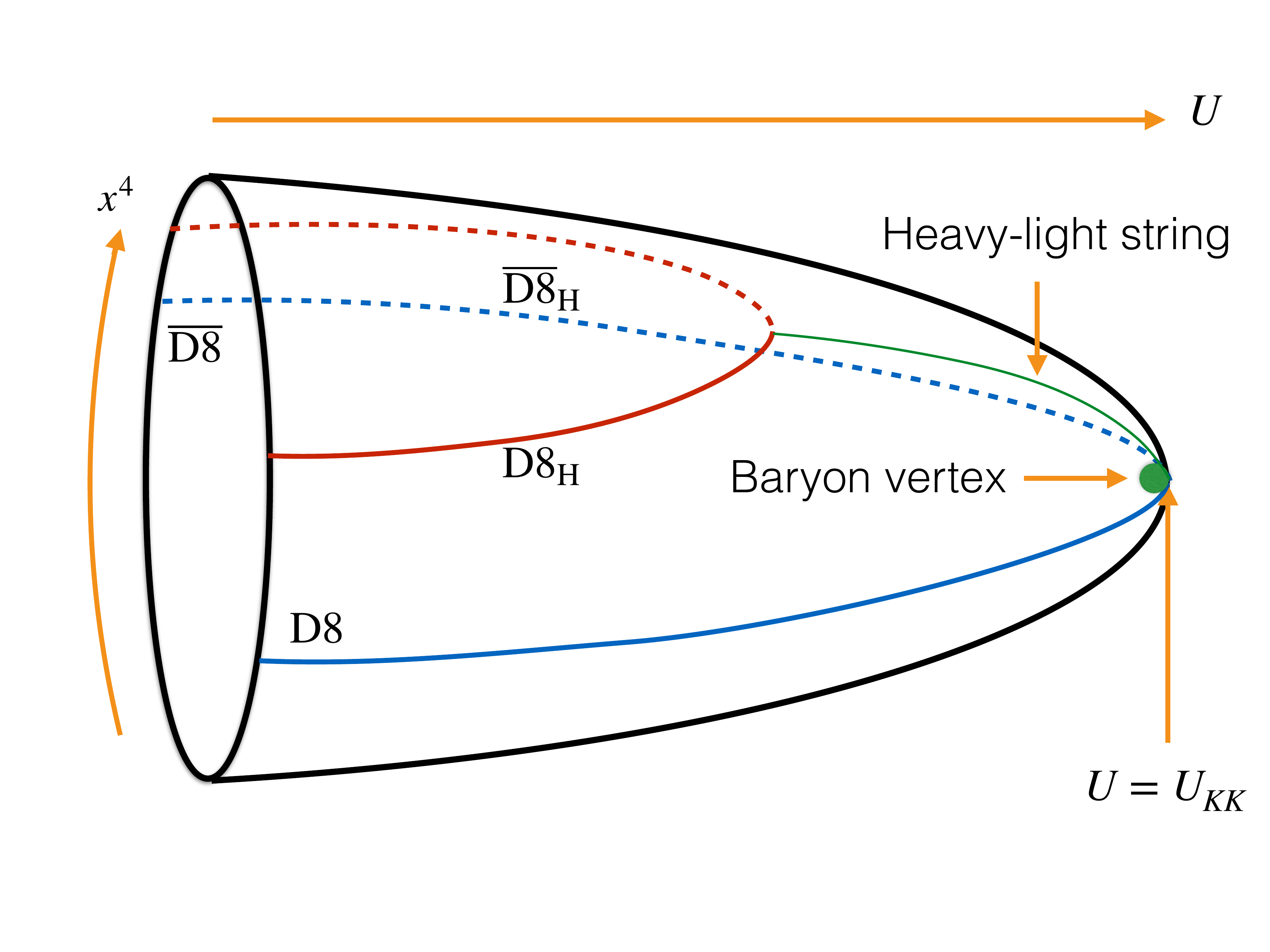}
\par\end{centering}
\caption{\label{fig:2}\textbf{ Left: }Higgs mechanism in string theory. A
stack of $N_{1}+N_{2}$ coincident D-branes move separately to become
two stacks of $N_{1}$ and $N_{2}$ coincident D-branes. The $U\left(N_{1}+N_{2}\right)$
gauge symmetry on the worldvolume breaks down into $U\left(N_{1}\right)\times U\left(N_{2}\right)$.
The multiplets produced by the open string becomes massive.\textbf{
Right: }Higgs mechanism in the D4-D8 model. The heavy-flavor $\mathrm{D8}/\overline{\mathrm{D8}}$-branes
are denoted by red. The heavy-light string is denoted by green. The
multiplets produced by the heavy-light string becomes massive thus
they can be identified as heavy-light mesons.}
\end{figure}

Due to the vanished minimized size of the 4-8 string\footnote{``4-8 string'' refers to the open string connecting D4- and D8-branes.},
the fundamental quark in the D4-D8 model is massless \cite{key-9}.
So it is very necessary to include the heavy flavor in the D4-D8 model
in order to describe quarks of heavy flavor. To achieve this goal,
let us follow the setup in \cite{key-50,key-51} in which the Higgs
mechanism in string theory is employed. Specifically, we can consider
the configuration of two stacks of the separated D-branes with an
open string connected them as it is illustrated in Figure \ref{fig:2}
. In this configuration, the $U\left(N_{1}+N_{2}\right)$ symmetry
on the worldvolume breaks down into $U\left(N_{1}\right)\times U\left(N_{2}\right)$
where $N_{1},N_{2}$ refers to the number of the coincident D-branes
in each stack. Thus the transverse modes of the D-brane acquire a
non-zero vacuum expectation value (VEV) which is recognized as the
separation of the D-branes. Therefore the multiplets produced by the
open string connected the separated D-branes will be massive due to
the VEV of the transverse modes \cite{key-52,key-53}, as the Higgs
mechanism in the standard model of the particle physics. 

Employing this configuration, another pair of $\mathrm{D8}/\overline{\mathrm{D8}}$-branes
as heavy flavor brane separated from the $N_{f}$ coincident $\mathrm{D8}/\overline{\mathrm{D8}}$-branes
with an open string (heavy-light string) stretched between them can
be introduced into the D4-D8 model as it is illustrated in Figure
\ref{fig:2}. Since the 8-8 string in the D4-D8 model is identified
as meson, the massive multiplets produced by the open string (heavy-light
string) stretched between the flavor branes is identified as the heavy-light
meson. Besides, as our concern is to include the heavy flavor in baryon,
we require that one endpoint of the heavy-light string is located
at $U=U_{KK}$ where the baryon vertex lives in this model. Afterwards,
the effective Lagrangian of the collective modes with heavy flavor
can be obtained by following the steps in Section 2. The notable derivation
here is the Yang-Mills gauge field and its field strength now becomes
an $\left(N_{f}+1\right)\times\left(N_{f}+1\right)$ matrix-valued
field as\footnote{The last element in $\boldsymbol{\mathrm{A}}_{\alpha}$ can be gauged
to be zero by the gauge symmetry.},

\begin{equation}
\mathcal{A}_{\alpha}\rightarrow\boldsymbol{\mathrm{A}}_{\alpha}=\left(\begin{array}{cc}
\mathcal{A}_{\alpha} & \Phi_{\alpha}\\
\Phi_{\alpha}^{\dagger} & 0
\end{array}\right),\ \mathcal{F}_{\alpha\beta}\rightarrow\boldsymbol{\mathrm{F}}_{\alpha\beta}=\left(\begin{array}{cc}
\mathcal{F}_{\alpha\beta}+i\alpha_{\alpha\beta} & f_{\alpha\beta}\\
f_{\alpha\beta}^{\dagger} & i\beta_{\alpha\beta}
\end{array}\right),\label{eq:65}
\end{equation}
where $\mathcal{A}_{\alpha},\mathcal{F}_{\alpha\beta}$ are $N_{f}\times N_{f}$
matrix-valued fields as we have specified in the previous sections.
$\Phi_{\alpha}$ is an $N_{f}\times1$ matrix-valued field which is
the multiplet created by the heavy-light string i.e. the heavy-light
meson field and\footnote{In our notation, the index in the square brackets is ranked as $T_{\left[\alpha\beta\right]}=\frac{1}{2!}\left(T_{\alpha\beta}-T_{\beta\alpha}\right)$.
And the gauge field is Hermitian $\boldsymbol{\mathrm{A}}_{\alpha}^{\dagger}=\boldsymbol{\mathrm{A}}_{\alpha}$. }

\begin{align}
\alpha_{\alpha\beta}= & 2\Phi_{[\alpha}\Phi_{\beta]}^{\dagger},\ \beta_{\alpha\beta}=2\Phi_{[\alpha}^{\dagger}\Phi_{\beta]},\nonumber \\
f_{\alpha\beta}= & 2\partial_{[\alpha}\Phi_{\beta]}+2iA_{[\alpha}\Phi_{\beta]}\equiv2D_{[\alpha}\Phi_{\beta]}.
\end{align}
For non-Abelian excitation on the D-brane, the standard DBI action
in (\ref{eq:4}) should include the dynamics of the transverse modes
of the D-brane, which is given as (up to quadratic term),

\begin{equation}
S\left[\varphi^{I}\right]=-T_{8}\frac{\left(2\pi\alpha^{\prime}\right)^{2}}{4}\int d^{9}x\sqrt{-\det g}e^{-\phi}\mathrm{Tr}\left\{ 2D_{\alpha}\varphi^{I}D_{\alpha}\varphi^{I}+\left[\varphi^{I},\varphi^{J}\right]^{2}\right\} ,\label{eq:67}
\end{equation}
where the index $I,J$ runs over the transverse space of the D-brane.
For the setup in the D4-D8 model with heavy flavor, $\varphi^{I}$
is an $\left(N_{f}+1\right)\times\left(N_{f}+1\right)$ matrix-valued
field with the covariant derivative $D_{\alpha}\varphi^{I}=\partial_{\alpha}\varphi^{I}+i\left[\mathbf{A}_{\alpha},\varphi^{I}\right]$
and the transverse coordinate of the D8-brane consists only of $x^{4}$
so that $\left(2\pi\alpha^{\prime}\right)\varphi^{I}\rightarrow x^{4}$
is the only T-dualitied transverse coordinate of the D8-brane. According
to \cite{key-52,key-53}, the moduli solution by the extrema of the
potential contribution can be given by $\left[x^{4},\left[x^{4},x^{4}\right]\right]=0$,
thus the moduli solution of $x^{4}$ for $N_{f}$ $\mathrm{D8}/\overline{\mathrm{D8}}$-branes
separated from one pair of heavy-flavored $\mathrm{D8}/\overline{\mathrm{D8}}$-branes
can be chosen as,

\begin{equation}
\frac{x^{4}}{2\pi l_{s}}=\left(\begin{array}{cc}
-\frac{v}{N_{f}}\boldsymbol{1}_{N_{f}} & 0\\
0 & v
\end{array}\right),\label{eq:68}
\end{equation}
where $v$ refers to the VEV of $x^{4}$, which is proportional to
the separation of the D8-branes in Figure \ref{fig:2}. Imposing (\ref{eq:68})
into (\ref{eq:67}), we can obtain a mass term for the heavy-light
field $\Phi_{\alpha}$ as \cite{key-54},

\begin{equation}
S\left[x^{4}\right]=-\tilde{T}v^{2}\frac{\left(N_{f}+1\right)^{2}}{N_{f}^{2}}\int d^{4}xdzU^{2}\left(z\right)\left(g^{zz}\Phi_{z}^{\dagger}\Phi_{z}+g^{\mu\nu}\Phi_{\mu}^{\dagger}\Phi_{\nu}\right),\label{eq:69}
\end{equation}
where $\tilde{T}=\frac{2}{3}T_{8}R^{3/2}U_{KK}^{1/2}\Omega_{4}g_{s}^{-1}$.
Therefore it is clear that if the heavy-flavored $\mathrm{D8}/\overline{\mathrm{D8}}$-brane
is coincident to the $N_{f}$ $\mathrm{D8}/\overline{\mathrm{D8}}$-branes
i.e. $v=0$, the heavy-light field $\Phi_{\alpha}$ becomes massless
so that the $U\left(N_{1}+N_{2}\right)$ symmetry becomes restored,
which means the action for $x^{4}$ (\ref{eq:67}) would be absent
in the DBI action given in (\ref{eq:4}), as it is in the original
model.

\subsection{Corrections to the heavy-light baryon spectrum}

Impose the replacement (\ref{eq:65}) into (\ref{eq:10}) and (\ref{eq:13}),
one can obtain the effective action for the heavy-light field $\Phi_{\alpha}$
in the large $\lambda$ limit as (up to quadratic order of $\Phi_{\alpha}$),

\begin{align}
\mathcal{L}_{H}\left[\Phi_{\alpha}\right]= & aN_{c}\lambda\mathcal{L}_{0}\left[\Phi_{\alpha}\right]+aN_{c}\mathcal{L}_{1}\left[\Phi_{\alpha}\right]+\mathcal{L}_{\mathrm{CS}}\left[\Phi_{\alpha}\right]+\mathcal{O}\left(\lambda^{-1}\right),\nonumber \\
\mathcal{L}_{0}\left[\Phi_{\alpha}\right]= & -\left(D_{M}\Phi_{N}-D_{N}\Phi_{M}\right)^{\dagger}\left(D_{M}\Phi_{N}-D_{N}\Phi_{M}\right)+2i\Phi_{M}^{\dagger}\mathcal{F}_{MN}\Phi_{N},\nonumber \\
\mathcal{L}_{1}\left[\Phi_{\alpha}\right]= & 2\left(D_{0}\Phi_{M}-D_{M}\Phi_{0}\right)^{\dagger}\left(D_{0}\Phi_{M}-D_{M}\Phi_{0}\right)-2i\Phi_{0}^{\dagger}\mathcal{F}^{0M}\Phi_{M}-2i\Phi_{M}^{\dagger}\mathcal{F}^{M0}\Phi_{0}\nonumber \\
 & +\frac{z^{2}}{3}\left(D_{i}\Phi_{j}-D_{j}\Phi_{i}\right)^{\dagger}\left(D_{i}\Phi_{j}-D_{j}\Phi_{i}\right)\nonumber \\
 & -2z^{2}\left(D_{i}\Phi_{z}-D_{z}\Phi_{i}\right)^{\dagger}\left(D_{i}\Phi_{z}-D_{z}\Phi_{i}\right)\nonumber \\
 & -\frac{2i}{3}z^{2}\Phi_{i}^{\dagger}\mathcal{F}_{ij}\Phi_{j}-2m_{H}^{2}\Phi_{M}^{\dagger}\Phi_{M},\label{eq:70}
\end{align}
and the CS term is

\begin{align}
\mathcal{L}_{\mathrm{CS}}\left[\Phi_{\alpha}\right]= & -\frac{N_{c}}{24\pi^{2}}\left(d\Phi^{\dagger}\mathcal{A}d\Phi+d\Phi^{\dagger}d\mathcal{A}\Phi+\Phi^{\dagger}d\mathcal{A}d\Phi\right)\nonumber \\
 & +\frac{iN_{c}}{16\pi^{2}}\left(d\Phi^{\dagger}\mathcal{A}^{2}\Phi+\Phi^{\dagger}\mathcal{A}^{2}d\Phi+\Phi^{\dagger}\mathcal{A}d\mathcal{A}\Phi+\Phi^{\dagger}d\mathcal{A}\mathcal{A}\Phi\right)\nonumber \\
 & +\frac{5N_{c}}{48\pi^{2}}\Phi^{\dagger}\mathcal{A}^{3}\Phi+\mathcal{O}\left(\Phi^{4},\mathcal{A}\right),
\end{align}
where the parameter $m_{H}$ is the energy scale of the heavy flavor
obtained by normalizing the mass term in (\ref{eq:70}) as $m_{H}=\frac{2}{3\sqrt{3}}\frac{N_{f}+1}{N_{f}}v$
and the associated equations of motion are obtained as,

\begin{align}
D_{M}D_{M}\Phi_{N}-D_{N}D_{M}\Phi_{M}+2i\mathcal{F}_{MN}\Phi_{M}+\mathcal{O}\left(\lambda^{-1}\right) & =0,\nonumber \\
D_{M}\left(D_{0}\Phi_{M}-D_{M}\Phi_{0}\right)-i\mathcal{F}^{0M}\Phi_{M}\nonumber \\
-\frac{1}{64\pi^{2}a}\epsilon_{MNPQ}\mathcal{K}_{MNPQ}+\mathcal{O}\left(\lambda^{-1}\right) & =0,
\end{align}
where

\begin{equation}
\mathcal{K}_{MNPQ}=i\partial_{M}\mathcal{A}_{N}\partial_{P}\Phi_{Q}-\mathcal{A}_{M}\mathcal{A}_{N}\partial_{P}\Phi_{Q}-\partial_{M}\mathcal{A}_{N}\mathcal{A}_{P}\Phi_{Q}-\frac{5i}{6}\mathcal{A}_{M}\mathcal{A}_{N}\mathcal{A}_{P}\Phi_{Q}.
\end{equation}
The above equation of motion refers to the static wave function of
heavy baryon which can be solved as $\Phi_{\alpha}=e^{\pm im_{H}t}\phi_{\alpha}\left(x\right)$
\footnote{The solution for $\Phi_{\alpha}$ may contain a contribution of $\mathcal{O}\left(\lambda^{-2/3}\right)$
when we use $\mathcal{A}=\mathcal{A}^{\mathrm{cl}}+\delta\mathcal{A}$.
So it has been neglected since our concern is the correction of $\mathcal{O}\left(\lambda^{-1/3}\right)$.} as,

\begin{align}
\phi_{0} & =-\frac{1}{1024a\pi^{2}}\left[\frac{25\rho}{2\left(x^{2}+\rho^{2}\right)^{5/2}}+\frac{7}{\rho\left(x^{2}+\rho^{2}\right)^{3/2}}\right]\chi,\nonumber \\
\phi_{M} & =\frac{\rho}{\left(x^{2}+\rho^{2}\right)^{3/2}}\sigma_{M}\chi,\label{eq:73}
\end{align}
where $\chi$ refers to the $SU\left(N_{f}\right)$ spinor independent
on $x$ and $\sigma_{M}$ is the embedded Pauli matrices as $\sigma_{M}/2=\left(t_{i},-\mathbf{1}_{N_{f}}\right)$.
Then follow the steps in Section 2 and \cite{key-19,key-22,key-23}
with our corrections of the BPST solution, we could take the limit
$m_{H}\rightarrow\infty$ to display mostly the contribution of the
heavy flavor and simplify the calculation. So in the double limit
$\lambda,m_{H}\rightarrow\infty$, the quantized Hamiltonian of the
collective modes with heavy flavor is finally calculated with (\ref{eq:73})
as,

\begin{align}
H_{\mathrm{HL}}= & H\left(\boldsymbol{\mathrm{K}}\right)+\left(N_{Q}-N_{\bar{Q}}\right)m_{H}+\Delta H+\mathcal{O}\left(\lambda^{-2/3}\right),\nonumber \\
\boldsymbol{\mathrm{K}}= & \frac{2N_{c}^{2}}{5}\left[1-\frac{5\sqrt{6}+10}{6}\frac{N_{Q}-N_{\bar{Q}}}{N_{c}}+\frac{65}{36}\frac{\left(N_{Q}-N_{\bar{Q}}\right)^{2}}{N_{c}^{2}}\right]-\frac{N_{c}^{2}}{3}\left(1-\frac{N_{Q}-N_{\bar{Q}}}{N_{c}}\right)^{2}\nonumber \\
 & +\frac{4}{3}\left(p^{2}+q^{2}+pq\right)+4\left(p+q\right)-2j\left(j+1\right),\label{eq:74}
\end{align}
where $N_{Q},N_{\bar{Q}}$, $H\left(\boldsymbol{\mathrm{K}}\right)$
refers respectively to the numbers of the heavy flavor, anti heavy
flavor and the Hamiltonian $H$ in (\ref{eq:51}) by replacing $K^{\prime}$
to $\boldsymbol{\mathrm{K}}$. Note that we have expressed all the
formulas in the unit of $M_{KK}$ which means $m_{H}$ has been rescaled
dimensionlessly as $m_{H}\rightarrow m_{H}M_{KK}$. Since the eigen
functions and spectrum can be obtained by replacing $K^{\prime}$
to $\boldsymbol{\mathrm{K}}$ in (\ref{eq:56}) (\ref{eq:59}), the
corrections to the heavy-light spectrum can also be evaluated by using
the standard method of quantum mechanics with $\Delta H$ as a perturbation. 

Keeping these in hand, let us attempt to fit the experimental data
of the heavy-light baryon. The lowest baryons with one heavy quark
are characterized by $n_{\rho}=0,1$, $N_{Q}=1,N_{\bar{Q}}=0$ and
$\left(p,q,j\right)=\left(0,1,0\right)$ for \textbf{$\bar{\mathbf{3}}$
}representation, $\left(p,q,j\right)=\left(2,0,1\right)$ for \textbf{$\mathbf{6}$
}representation due to their spin-1/2. The spin and parity of \textbf{$\bar{\mathbf{3}}$
}representation is $\frac{1}{2}^{+}$, so we can identify them as
$\Lambda,\Xi\left(\bar{\mathbf{3}}\right)$. The spin and parity of
\textbf{$\mathbf{6}$ }representation is $J=\frac{1}{2},\frac{3}{2}$,
so we can identify them as $\Sigma,\Xi\left(\mathbf{6}\right),\Omega$
or $\Sigma^{*},\Xi\left(\mathbf{6}\right),\Omega$. The parity of
the baryon state can be identified as $\left(-1\right)^{n_{Z}}$ corresponding
to the parity of the eigen function of $H_{Z}$ in the holographic
direction. Therefore, fitting the lowest \textbf{$\bar{\mathbf{3}}$
}representation by using the data of Particle Data Group $M_{\Lambda_{c}^{+}}^{\mathrm{exp}}\simeq2286\mathrm{MeV}$,
our calculation reveals the mass of the lowest \textbf{$\bar{\mathbf{3}}$}
and \textbf{$\mathbf{6}$} baryon with our corrections is very closed
to the experimental data as it is illustrated in Table \ref{tab:1}.
\begin{table}
\begin{centering}
\begin{tabular}{|c|c|c|c|c|c|}
\hline 
$\left(\mathrm{MeV}\right)$ & $\Lambda_{c}^{+}\left(\bar{\mathbf{3}}\right)$ & $\Xi_{c}^{+}\left(\bar{\mathbf{3}}\right)$ & $\Sigma_{c}^{+}\left(\mathbf{6}\right)$ & $\Xi_{c}^{'+}\left(\mathbf{6}\right)$ & $\Omega_{c}^{0}\left(\mathbf{6}\right)$\tabularnewline
\hline 
\hline 
$M$ & 2286 & 2451 & 2541 & 2567 & 2953\tabularnewline
\hline 
$M^{\mathrm{exp}}$ & 2286 & 2468 & 2453 & 2576 & 2697\tabularnewline
\hline 
\end{tabular}
\par\end{centering}
\caption{\label{tab:1} Mass spectrum of the lowest baryons with a single heavy
flavor $N_{Q}=1,N_{\bar{Q}}=0$. The value of $M$ is computed by
our spectrum while $M^{\mathrm{exp}}$ refers to the corresponding
experimental data. The parameter is set as $N_{c}=N_{f}=3$ for realistic
QCD and $M_{KK}=949\mathrm{MeV},\lambda=16.6$ as the meson data in
the D4-D8 model.}

\end{table}
 The notable point here is, we have employed again the meson data
in this model as $M_{KK}=949\mathrm{MeV},\lambda=16.6$ \cite{key-9,key-10},
which means the framework of the D4-D8 model can fit both the meson
and baryon spectra to the experimental data, thus it may become more
consistent with our corrections.

\section{The interaction of glueball and baryonic matters}

In this section, let us include the interaction of glueball and baryonic
matters in the D4-D8 model with our corrections. We first outline
the identification of glueball as the gravitational polarization in
the bulk, then specify the interaction of glueball and baryonic matters
with our corrections to the BPST solution.

\subsection{Glueball as the gravitational polarization}

According to gauge-gravity, the glueball field can be identified as
the gravitational fluctuation in the D4-D8 model \cite{key-25,key-26,key-27,key-28,key-29}.
The basic idea is that, as the bulk gravitational fluctuation is sourced
by the operators in the dual field theory and the background geometry
is produced by $N_{c}$ D4-branes as colors, so the mass spectrum
of the operators can be obtained by evaluating the pole of its correlation
functions. Since the bulk geometry is dual to the pure Yang-Mills
theory in holography, the operator, which sources the bulk gravitational
fluctuation, must relate to the energy-momentum tensor of Yang-Mills
theory thus it is gauge invariant. So this operator can be naturally
identified as glueball and its spectrum is therefore identified as
the mass of glueball in this model.

Recall the relation of the type IIA supergravity with $N_{c}$ D4-branes
solution and M-theory on $\mathrm{AdS_{7}}\times S^{4}$ \cite{key-4},
the generic formulas of the gravitational fluctuations in the D4-D8
model can be chosen as the 11d gravitational polarization on $\mathrm{AdS_{7}}$
given by \cite{key-26,key-27,key-28},

\begin{eqnarray}
\delta G_{44} & = & -\frac{r^{2}}{L^{2}}f\left(r\right)H_{G}\left(r\right)G\left(x\right),\nonumber \\
\delta G_{\mu\nu} & = & \frac{r^{2}}{L^{2}}H_{G}\left(r\right)\left[\frac{1}{4}\eta_{\mu\nu}-\left(\frac{1}{4}+\frac{3r_{KK}^{6}}{5r^{6}-2r_{KK}^{6}}\right)\frac{\partial_{\mu}\partial_{\nu}}{m_{G}^{2}}\right]G\left(x\right),\nonumber \\
\delta G_{11,11} & = & \frac{r^{2}}{4L^{2}}H_{G}\left(r\right)G\left(x\right),\nonumber \\
\delta G_{rr} & = & -\frac{L^{2}}{r^{2}}\frac{1}{f\left(r\right)}\frac{3r_{KK}^{6}}{5r^{6}-2r_{KK}^{6}}H_{G}\left(r\right)G\left(x\right),\nonumber \\
\delta G_{r\mu} & = & \frac{90r^{7}r_{KK}^{6}}{m_{G}^{2}L^{2}\left(5r^{6}-2r_{KK}^{6}\right)^{2}}H_{G}\left(r\right)\partial_{\mu}G\left(x\right),\label{eq:75}
\end{eqnarray}
where $x$ refers to the coordinates $x^{0,1,2,3}$ in 4d spacetime,
$r$ is the radial coordinate in the holographic direction, $m_{G}$
is the mass of the glueball. The 11d variables are related to the
type IIA supergravity solution (\ref{eq:1}) by,

\begin{equation}
L=2R,\ U=\frac{r^{2}}{2L},\ 1+z^{2}=\frac{r^{6}}{r_{KK}^{6}}=\frac{U^{3}}{U_{KK^{3}}}.\label{eq:76}
\end{equation}
Perform the dimension reduction, the 10d metric (\ref{eq:1}) involving
the 11d gravitational polarization (\ref{eq:75}) is given as,

\begin{align}
g_{\mu\nu} & =\frac{r^{3}}{L^{3}}\left[\left(1+\frac{L^{2}}{2r^{2}}\delta G_{11,11}\right)\eta_{\mu\nu}+\frac{L^{2}}{r^{2}}\delta G_{\mu\nu}\right],\nonumber \\
g_{44} & =\frac{r^{3}f}{L^{3}}\left[1+\frac{L^{2}}{2r^{2}}\delta G_{11,11}+\frac{L^{2}}{r^{2}f}\delta G_{44}\right],\nonumber \\
g_{rr} & =\frac{L}{rf}\left(1+\frac{L^{2}}{2r^{2}}\delta G_{11,11}+\frac{r^{2}f}{L^{2}}\delta G_{rr}\right),\nonumber \\
g_{r\mu} & =\frac{r}{L}\delta G_{r\mu},\ \ g_{\Omega\Omega}=\frac{r}{L}\left(\frac{L}{2}\right)^{2}\left(1+\frac{L^{2}}{2r^{2}}\delta G_{11,11}\right),\label{eq:77}
\end{align}
with the dilaton,

\begin{equation}
e^{4\phi/3}=g_{s}^{4/3}\frac{r^{2}}{L^{2}}\left(1+\frac{L^{2}}{r^{2}}\delta G_{11,11}\right).\label{eq:78}
\end{equation}
Here and $H_{G}\left(r\right)$ is determined by the eigen equation

\begin{equation}
\frac{1}{r^{3}}\frac{d}{dr}\left[r\left(r^{6}-r_{KK}^{6}\right)\frac{d}{dr}H_{G}\left(r\right)\right]+\left[\frac{432r^{2}r_{KK}^{12}}{\left(5r^{6}-2r_{KK}^{6}\right)^{2}}+L^{4}m_{G}^{2}\right]H_{G}\left(r\right)=0,\label{eq:79}
\end{equation}
where $m_{G}$ is the eigen value. Impose (\ref{eq:75}) and (\ref{eq:79})
to the 11d gravity action for $\mathrm{AdS_{7}}\times S^{4}$, we
can obtain

\begin{align}
S_{11\mathrm{D}}= & \frac{1}{2\kappa_{11}^{2}}\left(\frac{L}{2}\right)^{4}\Omega_{4}\int d^{7}x\sqrt{-\det G}\left(\mathcal{R}_{11\mathrm{D}}+\frac{30}{L^{2}}\right)\nonumber \\
= & \frac{1}{2}\int d^{4}x\left[\left(\partial_{\mu}G\right)^{2}+m_{G}^{2}G^{2}\right],\label{eq:80}
\end{align}
thus $G\left(x\right)$ can be identified as the massive scalar glueball
field in this model, specifically, it refers to the lowest state of
$J^{PC}=0^{++}$. The eigen equation is numerically solve in \cite{key-26,key-27,key-28}
which leads to the eigen values of the glueball in holography as it
is given in Table \ref{tab:2}.
\begin{table}
\begin{centering}
\begin{tabular}{|c|c|c|c|c|c|}
\hline 
Excitation & $n=0$ & $n=1$ & $n=2$ & $n=3$ & $n=4$\tabularnewline
\hline 
\hline 
Glueball mass $m_{G}$ & 0.901 & 2.285 & 3.240 & 4.149 & 5.041\tabularnewline
\hline 
\end{tabular}
\par\end{centering}
\caption{\label{tab:2} The glueball mass spectrum in the D4-D8 model in the
unit of $M_{KK}=1$.}

\end{table}

\subsection{Time-dependent perturbation for the collective modes}

In order to include the interaction of glueball and baryonic matters,
we need to derive Hamiltonian of the collective modes with the gravitational
fluctuations and the steps we should follow has been given in Section
2. Before this, we need to obtain the exact formula of the function
$H_{G}\left(r\right)$ which is determined by (\ref{eq:79}). Fortunately,
$H_{G}\left(r\right)$ can be solved analytically in the large $\lambda$
expansion satisfying,

\begin{equation}
H_{G}^{\prime\prime}\left(z\right)+\left(\frac{1}{z}+\frac{2z}{\lambda}\right)H_{G}^{\prime}\left(z\right)+\left(\frac{16}{3\lambda}+\frac{m_{G}^{2}}{\lambda}\right)H_{G}\left(z\right)+\mathcal{O}\left(\lambda^{-2}\right)=0,\label{eq:81}
\end{equation}
where we have expressed the formulas on $z$ coordinate (\ref{eq:7}),
imposed the rescaling (\ref{eq:9}) and rescaled $m_{G}$ dimensionlessly
as $m_{G}\rightarrow m_{G}M_{KK}$. Regularly, the solution to equation
(\ref{eq:81}) is obtained as a hypergeometric function. In the large
$\lambda$, we have

\begin{equation}
H_{G}\left(z\right)=\frac{\mathcal{C}}{M_{KK}}\left[1-\frac{16+3m_{G}}{12\lambda}z^{2}+\mathcal{O}\left(\lambda^{-2}\right)\right],\label{eq:82}
\end{equation}
where $\mathcal{C}$ is an integration constant. As a bulk fluctuation,
the constant $\mathcal{C}$ should satisfy $\mathcal{C}\ll1$. Picking
up the gravitational fluctuations (\ref{eq:77}) (\ref{eq:78}), in
the large $\lambda$ expansion, the dilaton and the inverse of the
induced metric on the D8-branes with gravitational fluctuations are
calculated with the rescaling (\ref{eq:9}) as (up to $\mathcal{O}\left(\lambda^{-1}\right)$),

\begin{align}
g^{\mu\nu}= & \frac{27}{8M_{KK}^{3}R^{3}}\left(1-\frac{z^{2}}{2\lambda}\right)\eta^{\mu\nu}+\frac{\mathcal{C}}{M_{KK}^{3}R^{3}}\bigg[\frac{135}{32m_{G}^{2}}\frac{\partial^{\mu}\partial^{\nu}G\left(x\right)}{M_{KK}^{2}}-\frac{81}{64}G\left(x\right)\eta^{\mu\nu}\nonumber \\
 & +\frac{27\left(22+3m_{G}\right)G\left(x\right)}{256}\frac{z^{2}}{\lambda}\eta^{\mu\nu}-\frac{45\left(38+3m_{G}\right)}{128m_{G}^{2}}\frac{\partial^{\mu}\partial^{\nu}G\left(x\right)}{M_{KK}^{2}}\frac{z^{2}}{\lambda}\bigg],\nonumber \\
g^{zz}= & \frac{27}{8M_{KK}R^{3}}\left(1+\frac{5z^{2}}{6\lambda}\right)+\frac{\mathcal{C}}{M_{KK}R^{3}}\left[\frac{189}{64}-\frac{9\left(202+21m_{G}\right)}{256}\frac{z^{2}}{\lambda}\right]G\left(x\right),\nonumber \\
g^{\mu z}= & -\frac{45}{4m_{G}^{2}M_{KK}^{2}R^{3}}\frac{\partial^{\mu}G\left(x\right)}{M_{KK}}\frac{z}{\sqrt{\lambda}}\mathcal{C},\nonumber \\
e^{-\phi}= & \frac{3}{8}\sqrt{\frac{3}{2}}\left(4-\frac{z^{2}}{\lambda}\right)g_{s}^{-1}M_{KK}^{3/2}R^{3/2}+\frac{3}{128}\sqrt{\frac{3}{2}}\left[-12+\frac{19+3m_{G}}{\lambda}z^{2}\right]g_{s}^{-1}M_{KK}^{3/2}R^{3/2},\label{eq:83}
\end{align}
where we have additionally rescaled $G\left(x\right)$ as $G\left(x\right)\rightarrow G\left(x\right)M_{KK}$
so that $G\left(x\right)$ is dimensionless glueball field in the
formulas. Note that $G\left(x\right)$ satisfies the equation of motion
from action (\ref{eq:80}) which accordingly refers to the wave function
for a free glueball as,

\begin{equation}
G\left(x\right)=\frac{1}{2}\left(e^{-ik_{\mu}x^{\mu}}+e^{ik_{\mu}x^{\mu}}\right),
\end{equation}
thus it should remain under the rescaling (\ref{eq:9}), and so does
$\partial_{\mu}G\left(x\right),\partial_{\mu}\partial_{\nu}G\left(x\right)$
since the derivatives relate to the momentum $k_{\mu}$ of the glueball. 

Then we insert the metric with the fluctuation (\ref{eq:77}) into
the DBI action (\ref{eq:4}) up to quadratic term as,

\begin{align}
S_{\mathrm{DBI}} & =-T_{8}\int_{\mathrm{D8}/\overline{\mathrm{D8}}}d^{9}xe^{-\phi}\mathrm{STr}\sqrt{-\det\left(g_{\alpha\beta}+2\pi\alpha^{\prime}\mathcal{F}_{\alpha\beta}\right)}\nonumber \\
 & =-T_{8}\Omega_{4}\int d^{5}xe^{-\phi}\sqrt{-\det g_{ab}}g_{\Omega\Omega}^{2}\left[1+\frac{1}{4}\left(2\pi\alpha^{\prime}\right)^{2}\mathcal{F}_{ab}\mathcal{F}^{ab}+...\right],
\end{align}
where the index $a,b$ runs over $0,1,2,3,z$. By imposing the rescaling
(\ref{eq:9}) , the effective Yang-Mills action is obtained as,

\begin{align}
S_{\mathrm{YM}}= & -\frac{1}{4}\left(2\pi\alpha^{\prime}\right)^{2}T_{8}\Omega_{4}\int d^{5}xe^{-\phi}\sqrt{-\det g_{ab}}g_{\Omega\Omega}^{2}\mathcal{F}_{ab}\mathcal{F}^{ab}\nonumber \\
\rightarrow & -\frac{1}{4}\left(2\pi\alpha^{\prime}\right)^{2}T_{8}\Omega_{4}\int d^{5}xe^{-\phi}\sqrt{-\det g_{ab}}g_{\Omega\Omega}^{2}\times\nonumber \\
 & \big(2\lambda^{-1}\mathcal{F}_{0M}\mathcal{F}_{0N}g^{00}g^{MN}+2\lambda^{-1}\mathcal{F}_{0N}\mathcal{F}_{M0}g^{0M}g^{N0}\nonumber \\
 & +4\lambda^{-1/2}\mathcal{F}_{0K}\mathcal{F}_{MN}g^{0M}g^{KN}+\mathcal{F}_{MN}\mathcal{F}_{KL}g^{MK}g^{NL}\big),\label{eq:86}
\end{align}
where the index $M,N,K,L$ runs over $1,2,3,z$. The metric presented
in (\ref{eq:86}) has been given in (\ref{eq:77}) and (\ref{eq:83}).
So while the kinetic terms for the collective modes remain as they
are given in Section 2, the potential for the collective modes depending
on the classical mass $M$ of the soliton would be obtained from the
onshell action (\ref{eq:86}) by recalling $S_{\mathrm{onshell}}=-\int M_{\mathrm{soliton}}dt$.
Note the Chern-Simons term in (\ref{eq:13}) is independent on the
metric, thus it is decoupled to to metric fluctuations (\ref{eq:75}).
Hence the classical soliton mass can be obtained after the metric,
dilaton with the fluctuations (\ref{eq:77}) (\ref{eq:83}) and the
instanton solution $\mathcal{A}^{\mathrm{cl}},\delta\mathcal{A}$
presented in Section 3 are all plugged into (\ref{eq:86}). Although
this calculation is very straightforward, the final result would be
tediously messy. In order to get a compact result, let us consider
the situation that the glueball is static or we are in the rest frame
of the glueball. In this sense, the momentum of the glueball becomes
$k_{\mu}=\left(m_{G},\mathbf{0}\right)$ (in the unit $M_{KK}=1$),
so that we have $\partial_{0}\rightarrow-im_{G},\partial_{i}\rightarrow0,G\left(x\right)=G\left(t\right)=\left(e^{-im_{G}t}+e^{im_{G}t}\right)/2$.
Then (\ref{eq:86}) can be simplified in the leading order of $\delta g_{ab}$,
as 

\[
S_{\mathrm{YM}}=S_{\mathrm{YM}}^{\mathrm{onshell}}+\delta S_{\mathrm{YM}}^{\mathrm{onshell}},
\]
where $S_{\mathrm{YM}}^{\mathrm{onshell}}$ refers to the action in
(\ref{eq:10}) and $\delta S_{\mathrm{YM}}^{\mathrm{onshell}}$ is
the leading-order coupling term of the gauge field and bulk fluctuations
calculated as,

\begin{align}
\delta S_{\mathrm{YM}}^{\mathrm{onshell}}= & \mathcal{C}\kappa\int d^{4}xdz\bigg\{\bigg[\frac{11}{32}\tilde{F}_{MN}^{2}-\frac{5}{4}\tilde{F}_{iz}^{2}+\frac{29}{16\lambda}\tilde{F}_{0M}^{2}-\frac{5}{4\lambda}\tilde{F}_{0i}^{2}\nonumber \\
 & +\left(\frac{89}{48}+\frac{9}{64}m_{G}\right)\frac{z^{2}}{\lambda}\tilde{F}_{iz}^{2}-\left(\frac{55}{96}+\frac{11}{128}m_{G}\right)\frac{z^{2}}{\lambda}\tilde{F}_{ij}^{2}+\frac{20z}{3m_{G}^{2}\lambda}F_{0i}F_{zi}\partial_{0}\bigg]G\left(t\right)\nonumber \\
 & +\mathcal{C}\kappa\left[\frac{9}{32\lambda}\left(\hat{F}_{0M}^{\mathrm{cl}}+\delta F_{0M}\right)^{2}+\frac{5}{8\lambda}\left(\hat{F}_{0z}^{\mathrm{cl}}+\delta F_{0z}\right)^{2}\right]G\left(t\right)\bigg\}.
\end{align}
After the integrating over $x^{i},z$ and using $\delta S_{\mathrm{YM}}^{\mathrm{onshell}}=-\int\Delta M_{L}dt$,
we can obtain a fluctuation of the soliton mass as,

\begin{align}
\Delta M_{L}= & -\mathcal{C}\kappa\pi^{2}G\left(t\right)\bigg\{\left[\frac{1}{2}+\frac{11}{24}\left(\frac{7}{6}\right)^{1/3}\left(\rho Z\right)^{4/3}\lambda^{-4/3}-\frac{3^{1/3}7^{2/3}}{2^{5/3}}\left(Z\rho\right)^{2/3}\lambda^{-2/3}\right]\nonumber \\
 & +\lambda^{-1}\left(\frac{17}{12}-\frac{1}{16}m_{G}\right)\left(2Z^{2}+\rho^{2}\right)+\frac{7}{320a^{2}\pi^{4}\rho^{2}\lambda}\bigg]\bigg\},
\end{align}
which implies a time-dependent term $\Delta H_{L}\left(t\right)=\Delta M_{L}\left(t\right)$
in Hamiltonian (\ref{eq:27}) would be presented when the bulk gravitational
fluctuations are taken into account. Here we use subscript ``$L$''
in $\Delta M_{L}\left(t\right)$ to refer to that there is no contribution
of heavy flavor to $\Delta M_{L}\left(t\right)$. As the bulk gravitational
fluctuations are identified as the glueball field, the interaction
of glueball and baryonic matters can be naturally included once the
time-dependent term $\Delta H_{L}\left(t\right)$ is added to (\ref{eq:27}).
And the decay of the baryonic matters involving glueball can be therefore
evaluated with this time-dependent Hamiltonian.

Besides, we can further include the contributions of heavy flavor
by using the replacement (\ref{eq:65}). In this sense, taking the
the double limit $\lambda,m_{H}\rightarrow\infty$, the variation
$\delta S_{\mathrm{YM}}^{\mathrm{onshell}}$ corresponding to the
fluctuation of the soliton is calculated as,

\begin{align}
\delta S_{\mathrm{YM}}^{\mathrm{onshell}} & =-\int dt\left[\Delta M_{L}+\Delta M_{H}+\mathcal{O}\left(m_{H}^{0}\right)\right],\nonumber \\
\Delta M_{H}\left(t\right) & =\frac{5}{2\lambda}\pi^{2}\kappa m_{H}^{2}G\left(t\right)\left(1-\frac{1}{8m_{H}a\pi^{2}\rho^{2}}\right)\mathcal{C},
\end{align}
where we have used subscript ``$H$'' in $\Delta M_{H}\left(t\right)$
to refer to the contribution of heavy flavor. Since the metric fluctuation
(\ref{eq:75}) is taken into account, there is another contribution
to the fluctuation of the soliton mass which comes from the action
(\ref{eq:67}) for the transverse modes of the D8-brane as,

\begin{align}
\delta S\left[x_{4}\right]= & -T_{8}\frac{\left(2\pi\alpha^{\prime}\right)^{2}}{4}\Omega_{4}\int d^{5}x\sqrt{-\det g_{\mathrm{D8}}}e^{-\phi}\left(\delta g^{\alpha\beta}-\delta\phi g^{\alpha\beta}\right)\Phi_{\alpha}^{\dagger}\Phi_{\beta}\nonumber \\
= & -\frac{16}{27}m_{H}^{2}\kappa\lambda^{-1}\mathcal{C}\int d^{4}xdz\bigg[-\frac{81}{64}+\frac{27\left(22+3m_{G}\right)}{256}\frac{z^{2}}{\lambda}\bigg]G\left(t\right)\delta^{ij}\phi_{i}^{\dagger}\phi_{j}+\nonumber \\
 & -\frac{16}{27}m_{H}^{2}\kappa\lambda^{-1}\mathcal{C}\int d^{4}xdz\left[\frac{189}{64}-\frac{9\left(202+21m_{G}\right)}{256}\frac{z^{2}}{\lambda}\right]G\left(t\right)\phi_{z}^{\dagger}\phi_{z},
\end{align}
by picking up (\ref{eq:73}). So the mass fluctuation from $\delta S\left[x_{4}\right]=-\int\Delta M_{x^{4}}dt$
is exactly computed with (\ref{eq:83}) as,

\begin{equation}
\Delta M_{x^{4}}\left(t\right)=-\frac{5}{2\lambda}\pi^{2}\kappa m_{H}^{2}G\left(t\right)\mathcal{C}+\mathcal{O}\left(\lambda^{-1}\right).
\end{equation}
Thus the total contribution $\Delta M_{HL}\left(t\right)$ involving
heavy flavor to the mass fluctuation is

\begin{equation}
\Delta M_{HL}\left(t\right)=\Delta M_{x^{4}}\left(t\right)+\Delta M_{H}\left(t\right)=-\frac{5}{16a\rho^{2}}m_{H}\kappa\lambda^{-1}G\left(t\right).
\end{equation}
Keeping the Hamiltonian for baryonic matters (\ref{eq:27}) with our
corrections (\ref{eq:49}) (\ref{eq:74}) in hand, we can compute
the transition amplitude with the time-dependent perturbed Hamiltonian
\begin{equation}
\Delta H\left(t\right)=\Delta M_{L}\left(t\right)+\Delta M_{HL}\left(t\right).\label{eq:93}
\end{equation}
 in our system in order to evaluate the decay of the baryon involving
glueball.

\subsection{Decay of baryonic meson involving the glueball}

In this section, let us evaluate the decay of the baryonic matters
involving the glueball quantitatively with this model. To begin with,
in experiment, there are some evidences that glueball may form in
the decays of some heavy-light mesons \cite{key-55,key-56} which
behaves like a baryon. Accordingly, let us consider $N_{f}=2$ for
the case of baryonic heavy-light meson (with no anti-heavy flavor).
So the Hamiltonian in (\ref{eq:74}) involving one heavy flavor becomes,

\begin{align}
H_{\mathrm{HL}}= & H^{N_{f}=2}\left(\boldsymbol{\mathrm{K}}\right)+\left(N_{Q}-N_{\bar{Q}}\right)m_{H}+\Delta H+\mathcal{O}\left(\lambda^{-2/3}\right),\nonumber \\
H^{N_{f}=2}\left(\boldsymbol{\mathrm{K}}\right)= & M_{0}+H_{\rho}^{N_{f}=2}\left(Q\right)+H_{Z}+\Delta H,\nonumber \\
H_{\rho}^{N_{f}=2}\left(Q\right)= & -\frac{1}{2m_{\rho}}\left[\frac{1}{\rho^{3}}\partial\left(\rho^{3}\partial_{\rho}\right)+\frac{1}{\rho^{2}}\left(\nabla_{S^{3}}^{2}-2Q\right)\right]+\frac{1}{2}m_{\rho}\omega_{\rho}^{2}\rho^{2},\nonumber \\
H_{Z}= & -\frac{1}{2m_{Z}}\partial_{Z}^{2}+\frac{1}{2}m_{Z}\omega_{Z}^{2}Z^{2},\nonumber \\
\Delta H= & -2\pi^{2}\kappa\lambda^{-4/3}\left(\frac{7}{6}\right)^{1/3}\left(\rho Z\right)^{4/3},
\end{align}
where

\begin{equation}
Q=\frac{N_{c}}{40\pi^{2}a}+\frac{N_{Q}}{8\pi^{2}a}\left(\frac{N_{Q}}{3N_{c}}-\frac{3}{4}\right).
\end{equation}
The eigen functions and energy spectrum of $H_{\rho}^{N_{f}=2}\left(Q\right)$
can be solve as

\begin{align}
\psi\left(\rho\right) & =e^{-\frac{m_{\rho}\omega_{\rho}}{2}\rho^{2}}\rho^{\tilde{l}}F\left(-n_{\rho},\tilde{l}+2;m_{\rho}\omega_{\rho}\rho^{2}\right)T^{\left(l\right)}\left(S^{3}\right),\nonumber \\
E\left(l,n_{\rho},n_{Z}\right) & =8\pi^{2}\kappa+\sqrt{\frac{\left(l+1\right)^{2}}{6}+\frac{640}{3}a^{2}\pi^{2}Q^{2}}+\frac{2\left(n_{\rho}+n_{Z}\right)+2}{\sqrt{6}},
\end{align}
where $F\left(-n_{\rho},\tilde{l}+2;m_{\rho}\omega_{\rho}\rho^{2}\right)$
is the hypergeometrical function, $T^{\left(l\right)}\left(S^{3}\right)$
is the Spherical harmonic function on $S^{3}$ and $\tilde{l}=-1+\sqrt{\left(l+1\right)^{2}+2m_{\rho}Q}$.
Note that $l$ is the quantum number of the angular momentum. The
eigen functions and energy spectrum of $H^{N_{f}=2}\left(\boldsymbol{\mathrm{K}}\right)$
can be obtained approximately by using $\Delta H$ as perturbation.
Afterwards, the decay rate of baryonic matters can be obtained by
using the standard technique in quantum mechanics as,

\begin{equation}
\Gamma_{i\rightarrow f}=\left|\left\langle f\left|\Delta H\left(t\right)\right|i\right\rangle \right|^{2}\delta\left(E_{f}-E_{i}-m_{G}\right).\label{eq:97}
\end{equation}
with the time-dependent term (\ref{eq:93}) in which the glueball
field is involved.

To close this section, let us attempt to fit the parameters to the
realistic QCD with $N_{c}=3$. For the baryonic meson, we set $N_{Q}=1,N_{f}=2,l=0,1,n_{Z}=1,3,5...$
due to $J^{P}=0^{-},1^{-}$ of the heavy-light meson. Then the mass
difference of the lowest heavy-light meson states with distinct angular
momentum is evaluated with our corrections as ($n_{\rho}=0,n_{Z}=1$),

\begin{equation}
M^{l=1}-M^{l=0}=0.171M_{KK}=162\mathrm{MeV},\label{eq:98}
\end{equation}
where the meson data $M_{KK}=949\mathrm{MeV},\lambda=16.6$ is also
picked up. In experiment, the lowest heavy-light meson states with
distinct angular momentum are $D^{*0},D^{0}$ whose mass difference
is

\begin{equation}
M_{D^{*0}}-M_{D^{0}}=141\mathrm{MeV},
\end{equation}
which is close to our (\ref{eq:98}). Besides, we find the various
decay processes among the lowest baryonic meson states ($l=0,n_{Z}=1$),
e.g. 

\begin{align}
1,\left|n_{\rho}=3\right\rangle  & \rightarrow\left|n_{\rho}=1\right\rangle +\left|m_{G}^{\left(n=0\right)}\right\rangle ,\nonumber \\
2,\left|n_{\rho}=5\right\rangle  & \rightarrow\left|n_{\rho}=0\right\rangle +\left|m_{G}^{\left(n=1\right)}\right\rangle ,
\end{align}
satisfy the constraint (\ref{eq:97}) and the associated decay rates
are computed in the limit $m_{H}\rightarrow\infty$ as,

\begin{align}
\Gamma_{1}/M_{KK} & =0.008\mathcal{C}m_{H}^{2},\nonumber \\
\Gamma_{2}/M_{KK} & =0.003\mathcal{C}m_{H}^{2}.
\end{align}
The parameter $m_{H}$ can be chosen as $m_{H}=0.129$ in order to
fit the mass of $D^{*0}$ in the heavy-light meson spectrum and the
value of constant $\mathcal{C}$ can be chosen as it is suggested
in \cite{key-26} i.e. $\mathcal{C}=144.545$ for $\left|m_{G}^{\left(n=0\right)}\right\rangle $;
$\mathcal{C}=114.871$ for $\left|m_{G}^{\left(n=1\right)}\right\rangle $.
In this sense, the lowest decay rates can be evaluated as $\Gamma_{1}=0.002M_{KK},\Gamma_{1}=0.004M_{KK}$.
Altogether, we are able to describe the decay of heavy-light meson
involving glueball in this holographic model while the exact property
of glueball is less clear in experiment.

\section{Summary and discussion}

In this work, we first derive the $\mathcal{O}\left(\lambda^{-1/3}\right)$
corrections to the BPST instanton solution on the flavor brane in
the D4-D8 model, which is a generalization of the $SU\left(2\right)$
case in \cite{key-24} in the strong coupling limit. The corrections
are obtained by solving the equations of motion for the gauge field
on the D8-branes with the same gauge condition for $\mathcal{A}^{\mathrm{cl}}$,
and minimizing the classical soliton mass. Then keeping our corrections
in hand, we follow \cite{key-19,key-20} in order to obtain the Hamiltonian
of collective modes, which describes the excitation of baryon. Afterwards,
the baryon states and spectrum are computed by solving the eigen equation
of the Hamiltonian of the collective modes according to the gauge-gravity
duality in this model. As the D4-D8 model is able to fit the meson
spectrum on the other hand, we therefore employ the meson data in
this model (i.e. the value for the unit $M_{KK}$ and t' Hooft coupling
constant $\lambda$ is set as $M_{KK}=949\mathrm{MeV},\lambda=16.6$
which are used to match the lowest meson spectrum) to fit the realistic
baryon spectrum in QCD with $N_{c}=3,N_{f}=3$. Using the standard
technique in quantum mechanics, we compute approximately the baryon
spectrum with our corrections which is very close to the experimental
data. Furthermore, the corrections to the heavy-light flavored baryon,
in which the heavy flavor is introduced by employing the Higgs mechanism
in string theory, is also taken into account in this work. So follow
the same steps to obtain the Hamiltonian of the collective modes,
we get the heavy-light baryon spectrum with our corrections and it
matches very well to the experimental data with the same value of
$M_{KK},\lambda$. Besides, we finally display how to include the
interaction of baryonic matter and glueball with our corrections.
As the glueball is identified as the bulk gravitational polarization
in this model, we obtain a fluctuation of the soliton mass due to
the bulk gravitational polarization and correspondingly a time-dependent
term arises in the Hamiltonian of the collective modes. Thus using
the standard method for time-dependent Hamiltonian in quantum mechanics,
it is possible to evaluate the decay rate of the baryonic matter involving
the glueball. Accordingly, we consider the $N_{f}=2$ heavy-light
meson as the baryonic matter and evaluate the decay rate caused by
the glueball field. Although the quantum mechanical description of
the baryonic matter decay involving glueball is natural and simply
in this model, the property of glueball is less clear in the current
experiment so that we do not attempt to further fit the experimental
data in this sector.

The remarkable point of this work is that, by picking up our corrections,
it is possible to fit the lowest spectrum of two-flavor light meson,
three-flavor baryon and two-flavor heavy-light meson with same meson
data i.e. $M_{KK}=949\mathrm{MeV},\lambda=16.6$ which are the only
parameters in our theory. In this sense, this work is a good improvement
of the D4-D8 framework and \cite{key-19,key-20,key-22,key-23} (In
\cite{key-19,key-20,key-22,key-23}, the infinitely large t' Hooft
coupling constant $\lambda$ is strictly necessary in theory thus
it is unable to employ the meson data of $M_{KK},\lambda$ in the
D4-D8 model). In addition, this work also introduces the next leading
order correction to the baryon vertex through the instanton configuration
in the large $\lambda$ expansion, which is to equivalently consider
the leading order interaction among the instantons. Therefore the
instanton configuration with our corrections may be more close to
the reality when they are employed to investigate the other features
of baryonic matter such as its phase diagrams as \cite{key-11,key-57}.
And we will leave this part for the future work.

\section*{Acknowledgements}

This work is supported by the National Natural Science Foundation
of China (NSFC) under Grant No. 12005033, the research startup foundation
of Dalian Maritime University in 2019 under Grant No. 02502608 and
the Fundamental Research Funds for the Central Universities under
Grant No. 3132022198.

\end{document}